\documentclass[a4paper,fleqn,usenatbib]{mn2e}
\usepackage[T1]{fontenc}
\usepackage{ae,aecompl}
\long\def\symbolfootnote[#1]#2{\begingroup%
\def\thefootnote{\fnsymbol{footnote}}\footnote[#1]{#2}\endgroup}

\usepackage{rotating}
\usepackage{graphicx}	

\newcommand{\msun}{\mbox{\rm M$_{\odot}$}}
\usepackage{amsmath}
\usepackage{multicol}

\newcommand{\arcs}{\hbox{$^{\prime\prime}$}}
\newcommand{\arcm}{\mbox{$^{\prime}$}}

\newcommand{\ra}{\mbox{$\alpha_{2000}$}}
\newcommand{\dec}{\mbox{$\delta_{2000}$}}
\newcommand{\jhk}{\mbox{$JHK_{\rm s}$}}
\newcommand{\ks}{\mbox{$K_{\rm s}$}}

\newcommand{\degree}{\mbox{$^{\circ}$}}
\newcommand{\av}{\mbox{$A_{\rm V}$~}}

\newcommand{\hii}{\mbox{H{\sc~ii}~}}

\newcommand{\mum}{\hbox{$\mu$m}~}

\def\halpha{\mbox{H$\alpha$}\,\,}

\def\farcs{\hbox{$.\!\!^{\prime\prime}$}}

\def\av{\mbox{$A_{_V }$}}

\def\gsim{\;\rlap{\lower 2.5pt\hbox{$\sim$}}\raise 1.5pt\hbox{$>$}\;}
\def\lsim{\;\rlap{\lower 2.5pt\hbo time lag between the formation x{$\sim$}}\raise 1.5pt\hbox{$<$}\;}
\def\la{\mathrel{\hbox{\rlap{\hbox{\lower4pt\hbox{$\sim$}}}\hbox{$<$}}}}
\def\ga{\mathrel{\hbox{\rlap{\hbox{\lower4pt\hbox{$\sim$}}}\hbox{$>$}}}}

\def\fdg{\hbox{$.\!\!^\circ$}}
\def\farcm{\hbox{$.\mkern-4mu^\prime$}}
\def\farcs{\hbox{$.\!\!^{\prime\prime}$}}

\def\vi{\hbox{$(V\!-\!I)$ }}               
\def\jhk{\hbox{$J\!H\!K_{\rm s}$}~}              
\def\jh{\hbox{$(J\!-\!H)$ }}               
\newcommand{\hk}{\hbox{$(H\!-\!K_{\rm s})$}~}
\newcommand{\jk}{\hbox{$(J\!-\!K_{\rm s})$}~}

\usepackage{longtable}
\usepackage{multicol}
\usepackage{pdflscape}
\usepackage{amssymb}
\usepackage{xcolor}
\usepackage{psfrag}
\usepackage{soul}
\usepackage{tikz}
\usepackage[colorlinks=true,citecolor=blue]{hyperref}

\title[Young stellar population and star formation activities in the S311 region]{A multiwavelength investigation of the \hii region S311: Young stellar population and star formation}
\author[Ram Kesh Yadav et al.]
{Ram Kesh Yadav,$^{1,2}$\thanks{E-mail: ram$\_$kesh@narit.or.th} 
A. K. Pandey,$^{2}$ 
Saurabh Sharma,$^2$ 
D. K. Ojha,$^3$ 
M. R. Samal,$^4$ \and 
K. K. Mallick,$^3$ 
J. Jose,$^5$ 
K. Ogura,$^6$ 
Andrea Richichi,$^1$ 
Puji Irawati,$^1$ 
N. Kobayashi,$^7$ \and 
and C. Eswaraiah$^8$ \\
$^{1}$National Astronomical Research Institute of Thailand, Chiang Mai, 50200, Thailand\\
$^{2}$Aryabhatta Research Institute of Observational Sciences, Nainital 263001, India\\
$^{3}$Tata Institute of Fundamental Research, Homi Bhabha Road, Mumbai-400005, India\\
$^{4}$Graduate Institute of Astronomy, National Central University 300, Jhongli City, Taoyuan County - 32001, Taiwan\\
$^{5}$Kavli Institute for Astronomy and Astrophysics, Peking University, 5 Yiheyuan Road, Haidian District, Beijing 100871, P. R. China\\
$^{6}$Kokugakuin University, Higashi, Shibuya-ku, Tokyo - 1508440, Japan\\
$^{7}$Kiso Observatory, School of Science, University of Tokyo, Mitake, Kiso-machi, Kiso-gun, Nagano-ken 397-0101, Japan\\
$^{8}$National Tsing Hua University, Hsinchu 30013, Taiwan}

\begin{document}

\date{}

\maketitle

\label{firstpage}

\begin{abstract}
 We present a multiwavelength investigation of the young stellar population and star formation activities around the \hii region Sharpless 311. 
Using our deep near-infrared observations and archival {\it Spitzer}-IRAC observations, we have detected a total of 125 young stellar objects (YSOs) in an area of $\sim$86 arcmin$^2$. The YSOs sample include 8 Class I and 117 Class II candidate YSOs. The mass completeness of the identified YSOs sample is estimated to be 1.0 \msun. The ages and masses of the majority of the  candidate YSOs are estimated to be in the range of $\sim$0.1$-$5 Myr and $\sim$0.3$-$6 \msun, respectively. The 8 \mum image of S311 displays an approximately spherical cavity around the ionizing source which is possibly created due to the expansion of the \hii region. The spatial distribution of the candidate YSOs reveals that a significant number of them are distributed systematically along the 8 $\mu$m emission with a majority clustered around the eastern border of the \hii region. Four clumps/compact \hii regions are detected in the radio continuum observations at 1280 MHz, which might have been formed during the expansion of the \hii region. The estimated dynamical age of the region, main-sequence lifetime of the ionizing source, the spatial distribution and ages of the candidate YSOs indicate triggered star formation in the complex.
\end{abstract}

\begin{keywords}
Stars: clusters -- star clusters: individual (Haffner 18) -- stars: formation -- stars: pre-main sequence -- ISM: \hii regions -- ISM: Individual (S311)
\end{keywords}

\section{Introduction}
In recent years \hii regions have been studied quite extensively on account of their close association with star formation. There seems to be two  major modes of star formation associated with \hii regions depending on the initial density distribution of the natal molecular cloud. One is the clustered mode which gives birth to a  more or less rich open cluster and the other is the dispersed mode which forms only loose clusters or aggregates of stars. Presumably, the former takes place in centrally condensed, massive clouds, whereas the latter occurs in clumpy, dispersed clouds \citep[see e.g.,][]{Ogura2006BASI...34..111O}. These clusters/aggregates of stars emerging from their natal clouds are useful laboratories to address some of the fundamental questions of star formation.

Massive stars in a star-forming region (SFR) are believed to have a significant effect on the entire region. Their energetic ultra-violet (UV) radiation and stellar wind as well as blast waves from a supernova explosion can evaporate or disperse the surrounding cloud and consequently may terminate star formation in the region. Alternatively, they may play a constructive role in inciting a sequence of star formation activities in the neighbourhood. The interaction of an \hii region with the molecular cloud can trigger star formation of the next generation through two main processes, namely `collect and collapse (CC)' and `radiation driven implosion (RDI)'. In the CC process, a compressed layer of neutral material is accumulated between the ionization front (IF) and the shock front (SF) around the expanding \hii region. The dynamical instabilities in the high density layer result in fragmentation of the molecular gas and formation of second-generation stars including massive stars \citep{Elmegreen1977ApJ...214..725E}. Recent simulations of this process are shown by \citet{Hosokawa2005ApJ...623..917H,Hosokawa2006ApJ...646..240H} and    \citet{Dale2007MNRAS.375.1291D}. An observational signature of this process is the presence of a dense layer and (more or less regularly spaced) massive condensations adjacent to the \hii region \citep[e.g.,][]{Deharveng2003a,Deharveng2005,Zavagno2006A,Liu2015,Pirogov2013,Ohlendorf2013}. If the IF/SF encounters pre-existing denser region, it compresses them to induce star formation. This process is called RDI \citep[e.g.,][]{Bertoldi1989ApJ...346..735B,Lefloch1995A&A...301..522L}. The signatures of the RDI process are the anisotropic density distribution in a relatively small molecular cloud surrounded by a curved IF (bright rim) as well as a group of young stellar objects (YSOs) aligned from the bright rim towards the direction of the ionizing source(s). Several authors \citep[e.g.,][]  {Bertoldi1989ApJ...346..735B,Lefloch1995A&A...301..522L,Lefloch1997A&A...324..249L,De2002ApJ...577..798D,Kessel-Deynet2003MNRAS.338..545K,Miao2006MNRAS.369..143M}  have carried out detailed model calculations of the RDI process. The distribution of YSOs and the morphological details of the environment around massive star(s) are frequently used to probe the star formation history of the region \citep[e.g.,][and references therein]{Ogura2007,Chauhan2009MNRAS.396..964C,Pandey2013ApJ...764..172P, Pandey2013NewA...19....1P}.

\section{The \hii region S311}
\begin{figure*}
\centering
\includegraphics[width=\columnwidth]{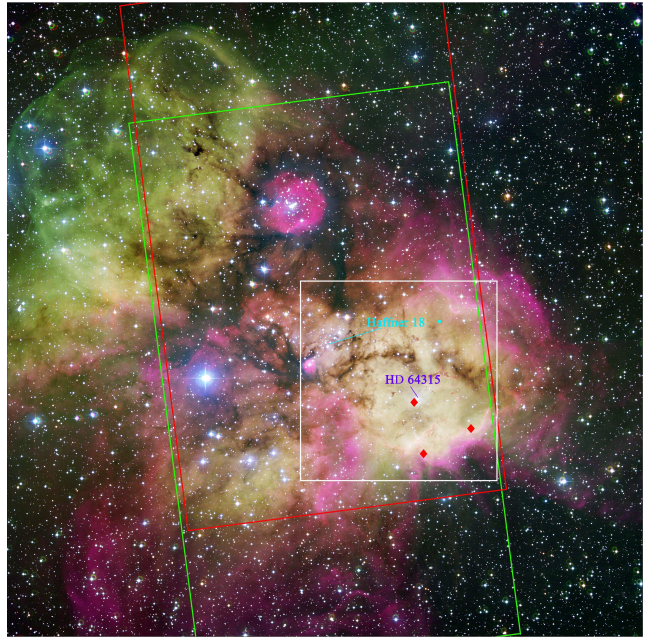}
\caption{\label{eso_im} The ESO Wide Field Imager (WFI) optical colour-composite image (ESO photo release eso0544) made using $B$ band in blue, $V$ band in green, $R$ band in red, OIII emission in green, and \halpha in red. The white box marks the $\sim$9.2$\times$9.4 arcmin$^2$ area (same as Fig. 2) used in present study. The area covered by the green and red lines represent the approximate areas observed by IRAC 3.6 and 4.5 $\mu$m bands, respectively. The location of main ionizing source (HD 46315) of S311 is marked. The diamonds mark the location of the IRAS sources. The open cluster Haffner 18 (H18) is located at the northeastern border of the white box. North is up, and east is to the left.} 
\end{figure*}

 The \hii region Sharpless 311 \citep[][hearafter S311]{Sharp1953ApJ...118..362S}, also known as NGC 2467 (\ra = 07$^{\rm h}$52$^{\rm m}$26$^{\rm s}$, \dec = -26\degree26\arcm12\arcs; $l$ = 243\fdg2, $b$ = 0\fdg4), is a part of the Puppis OB association. Fig. \ref{eso_im} shows the ESO Wide Field Imager (WFI) optical colour-composite image of S311.  The complex is interesting as it apparently contains two young open clusters Haffner 18 (H18) and Haffner 19 (H19, is located towards north-eastern direction of H18 in Fig. \ref{eso_im}). The most interesting structure is, S311 which is excited by an O6V type star HD 64315 \citep{Crampton1971,Keenan1980ApJS...42..541K,Walborn1982}. The location of three IRAS sources are also shown with diamond symbols. The white square shows the region of interest in the present study. Fig. \ref{eso_im} also delineates several dark and bright ridges, finger-like features, a dark filamentary structure (possibly due to large extinction), and extended ionized emission. 

The cluster H18 in Fig \ref{eso_im}, apparently close to the central nebula, is elongated along the Galactic plane with several OB type stars \citep{Fitzgerald1974,Munari1998,Yadav2015,Moreno-Corral2005RMxAA..41...69M}. The entire region has many other early-type stars, e.g., HD 64568 (O3V), CD -26 5129 (O7), and CD -26 5126 (B0.5V), whose associations with the region are not clear \citep[see e.g.][]{Munari1996,Munari1998,Feinstein1989}.

In the past, several studies based on photometric, spectroscopic and kinematic observations yielded conflicting results on distance estimation. H18 is a confusing one as it is reported to be composed of two stellar groups, \citep[][]{Vazquez2010,Yadav2015}. The distance estimate to H18 varies from 4.1 to 11.2 kpc \citep{Vazquez2010,Yadav2015,Moreno-Corral2005RMxAA..41...69M}.  The photometric studies of the S311 region by \citet{Havlen1972, Georgelin1973, Fitzgerald1974,Feinstein1989} and \citet{Yadav2015} yield mean distances of 4.2, 4.2, 6.9, 4.1 and 5.0 kpc, respectively. The radial velocity observations of the region obtained  by \citet{Georgelin1973} and \citet{Pismis1976RMxAA...1..373P} estimated the kinematic distances to be 3.51 and 4.2 kpc, respectively, whereas \citet{Balser2011ApJ...738...27B} have estimated a kinematic distance of 5.3 kpc. The large scatter in the photometric distances is mainly due to different numbers and sets of stars used for the distance estimation. The kinematic distances differ mainly due to the use of different rotation curve models. Using the isochrone fitting method and spectroscopy of the ionizing source, \citet{Yadav2015} estimated the distance to the region as 5.0 kpc.  In the present study we adopt the photometric distance of 5.0 kpc which is also in agreement with the recent kinematic distance estimates by \citet{Balser2011ApJ...738...27B}.

\citet{snider2009ApJ...700..506S} have studied the region in an area covering approximately 400 arcmin$^2$ using shallow Two Micron All Sky Survey (2MASS) \citep{Skrutskie2006} in the near-infrared (NIR) and {\it Spitzer}  Space Telescope mid-infrared (MIR) observations and identified 45 candidate YSOs. They have estimated ages and masses of candidate YSOs on the basis of the spectral energy distribution (SED) analysis and found to be in the range of 0.002$-$1.2 Myr and 0.35$-$5.73 \msun, respectively. The total number of stars were estimated to be around 5000 by using the Salpeter IMF in an area of $\sim$400 arcmin$^2$ having masses in the range of 0.2$-$40 \msun. They also concluded that the spatial distribution of candidate YSOs in the region suggests an ongoing star formation which might have been triggered due to the expansion of the \hii  region.

Since low mass stars constitute the vast majority of the Galactic stellar population, their identification, characterization and distribution are essential to draw any inference on star formation history of a complex. Deep NIR and MIR photometry is an ideal tool to uncover the low mass stellar content in heavily obscured and dense environments. In the present study we focus on understanding the
 star formation scenario in the \hii S311 using the low mass stellar content in the region. We identify and characterize low mass YSOs in the S311 region on the basis of deep NIR observations carried out with the Cerro Tololo Inter-American Observatory (CTIO) 4-m Blanco telescope in conjugation with archival {\it Spitzer}-IRAC observations. The identified YSOs were used to understand star formation scenario of the complex. Giant Metrewave Radio Telescope (GMRT) radio continuum observations have been used to investigate the properties of the ionized gas. In Section \ref{obs_red} we describe the observations and data reduction.  
 Section \ref{ana_res} describes the identification of YSOs using NIR and MIR colours, extinction in the region, emission from ionized gas, age and mass estimates of YSOs, their spatial distribution, and mass distribution. In Section \ref{discu} we discuss the star formation history in the region and we conclude our study in Section \ref{concl}. 

\section{Observations and data reduction}
\label{obs_red}
\subsection{Near infrared photometry}
We observed the \hii region S311 (centered on \ra = 07$^{\rm h}$52$^{\rm m}$24$.\!\!^{\rm s}$22, \dec = -26\degree24\arcm58\farcs40)  in NIR broad-bands $J$ (1.25 \mum), $H$ (1.63 \mum) and {\ks~} (2.14 \mum) on 2010 March 3 using  the Infrared Side Port Imager (ISPI) camera \citep[][]{vander2004SPIE.5492.1582V} mounted on the CTIO Blanco 4-m telescope. The ISPI has a 2048 $\times$ 2048 HgCdTe HAWAII-2 array with a pixel scale of  0.3 arcsec pixel$^{-1}$, which gives a field of view (FoV) of 10.25 $\times$ 10.25 arcmin$^2$. Nine dithered frames, each having an offset of $\sim$30 arcsec were observed with an exposure of 60s each, to compensate for the cosmetic effects of the detector, and also to create a sky image for the sky subtraction. Thus the total integration time was 540s in each of the $J$, $H$ and {\ks~} bands. The sky conditions were good in general and the seeing during the observations was approximately 1.0 arcsec. Dark frames and dome flats were obtained at the beginning and end of the observations.

The reduction was done with IRAF following the standard procedure: dark subtraction, flat-fielding, sky subtraction, alignment and averaging of the  dithered frames. The offsets between the dithered frames were determined with respect to the reference frame. Thereafter, these dithered frames were mean combined and trimmed at the common area. After trimming, the final FoV of the \jhk images for photometry was found to be $\sim$9.2$\times$9.4 arcmin$^2$. The final frames were {\it wcs} updated by using ``koords'' task of ``karma'' using the 2MASS image of the same area as a reference.  The residual {\it rms} error of the ISPI plate solution with respect to 2MASS was 0.2 arcsec. Fig. \ref{nir_cc_im} shows the colour-composite image of the region made from the three final frames in  {\ks~} (red), $H$ (green) and $J$ (blue) bands. Photometry was performed by using the DAOPHOT-II package in IRAF {\citep{Stetson1987PASP...99..191S}. Since  the region is crowded  as well as has a large amount of diffuse emission, point-spread function (PSF) photometry was performed. The PSF was constructed by using 38, 29 and 26 well isolated bright stars in $J$, $H$ and \ks -bands, respectively, which were free from the nebulosity and distributed  throughout the frames. The PSF stars were identified visually in the interactive mode. The tasks ``PSTSELECT'' and ``PSF'' were used to select the  stars and construct the  PSF, respectively.  The photometry on the final images was carried out by using  ``DAOPHOT-II/ALLSTAR'' routines described in \citet{Stetson1987PASP...99..191S}. Our deep NIR photometry detected 3406, 3706 and 3831 point sources above 4$\sigma$ sky background in $J$, $H$, and \ks -bands, respectively. These sources in individual bands were merged (with a matching radius equal to the average seeing of 1 arcsec) to produce a  $J$, $H$,  {\ks~}  catalogue. In the case of multiple matches the nearest match was considered. We consider only those sources having error $<$ 0.1 mag in all the three bands, resulting a final catalogue of 2671 point sources. This catalogue is used to identify candidate YSOs having NIR-excesses (cf. Section \ref{nir_sec}) and to produce an extinction map (cf. Section \ref{redd_map}).

 Photometric calibration of the instrumental $J$, $H$, and {\ks~} magnitudes was done by using the 2MASS point sources in the FoV obtained from the Vizier catalogue service, as explained below. We obtained 314 2MASS sources with the constraints of ``phqual = AAA" and ``ccflag = 000" in the FoV. These constraints ensure highest photometric quality detection without neighbouring source confusion in all the three bands. We first constructed a NIR \jh/\hk  two-colour diagram (NIR-TCD) using these data to identify only the main-sequence (MS) stars in the region. The sources having $(J-H)\leq0.6$ mag (288 sources) were considered to be MS sources. These 288 MS stars, were then matched with ISPI data with a matching radius of 1.0 arcsec which yielded 182 common sources. These sources were then used to carry out the photometric calibration  by using the following relations:\\
\[(J-H)_{2MASS} =(1.020\pm0.027)\times(j-h)_{ISPI} +( 0.015\pm0.015)\]
\[(H-K)_{2MASS} =(0.957\pm0.028)\times(h-k)_{ISPI} +(-0.003\pm0.008)\]
\[(J-K)_{2MASS} =(0.948\pm0.050)\times(j-k)_{ISPI} +( 0.099\pm0.067)\]
\[K_{2MASS}       = k_{ISPI} + (-0.032\pm0.086)\times(J-K)_{2MASS}+( 0.007\pm0.070)\]
 Stellar sources brighter than 10 mag in all the three bands were found to be saturated and these saturated sources were replaced with the corresponding 2MASS magnitudes. The final calibrated data were then converted to the California Institute of Technology (CIT) system by using the colour transformation equations given by \citet{Carpenter2001AJ....121.2851C}. The final photometric catalogue of the NIR data containing 2671 sources has been used for further analyses and shown in Table \ref{phot_data}. 

\subsection{Optical slitless spectroscopy}\label{slitless}
  \halpha slitless spectroscopic observations of the S311 region were carried out by using the Himalayan Faint Object Spectrograph Camera (HFOSC) on the 2-m Himalayan Chandra Telescope (HCT) on 2007 November 16. The spectra were obtained by the combination of a wide \halpha  filter (6300-6740 {\AA}) and grism 5 (5200-10300 {\AA}) with a spectral resolution of 870 {\AA}. The central 2048 $\times$ 2048 pixels of the 2048 $\times$ 4096 CCD were used for the observations. Two slitless spectra each with an exposure time of 420 s were obtained and were coadded 
in order to increase the signal-to-noise (S/N) ratio. Emission line stars were visually identified as an enhancement of \halpha emission above the continuum. Two sources are found to be emission line stars.

\subsection{{\it Spitzer}$-$IRAC observations}\label{spit_obs}
 The {\it Spitzer}$-$IRAC observations for the S311 region (PID: 20726) were made on 2006 May 3 using the 3.6, 4.5, 5.8 and 8.0 \mum bands (hereafter ch1, ch2, ch3 and ch4) and were downloaded from the {\it Spitzer} heritage archive (SHA). The data were obtained in the high dynamic range mode with five dithered positions per map and with integration time of 0.4 and 10.4s per dither. The basic calibrated data images (version S18.18.0) were downloaded, processed and calibrated using the IRAC pipelines. The MOsaicker and Point source EXtractor package (MOPEX) pipeline (version 18.5.4) was used to create the final mosaics with an image scale of 1.2  arcsec pixel$^{-1}$. The area covered by the IRAC observations, displayed in Fig. \ref{eso_im}, is $\sim$350 arcmin$^2$. The ch1 and ch2 images have an spatial offset of $\sim$6.7 arcmin.

As the region is crowded and has strongly varying background, aperture photometry is not applicable. Therefore, we performed the point response function (PRF) fitting method in the multi-frame mode using the tool Astronomical Point source EXtractor (APEX) developed by the {\it Spitzer} Science Center on all the {\it Spitzer}-IRAC images to extract the magnitudes. The standard PRF map table\footnote{http://ssc.spitzer.caltech.edu/irac/calibrationfiles/psfprf/prfmap.tbl} provided on the {\it Spitzer} website was used to fit variable PRFs across the image. Point sources with peak values more than 5$\sigma$ above the background were considered for source detection. On visual examination, many sources detected in the nebulosity appeared to be spurious. These spurious sources were deleted from the automated detection list. Moreover, some of the sources were not detected by the APEX automatic routine. These sources were identified visually and added to the coordinate list in the user-list mode in APEX to extract the magnitudes and thus we made sure that photometry of every genuine source is derived. In ch1, 185 and 284 sources had to be removed and added, respectively, whereas in ch2, 298 and 83 sources had to be removed and added respectively. A common catalogue was made by using the sources detected in ch1 and ch2, and then was used as input list to detect sources in ch3 and ch4. The adopted zero-points for conversion between the flux densities and magnitudes were taken as 280.9, 179.7, 115.0 and 64.1 Jy in the ch1, ch2, ch3 and ch4, respectively, following the IRAC Data Handbook \citep{Reach2005}. The magnitudes of saturated bright sources in the long exposure frames were replaced with those from short exposure frames. Because ch1 and ch2 are more sensitive than ch3 and ch4 and are less affected by the bright diffuse emission that dominates in the ch3 and ch4 images, ch3 and ch4 detected far fewer sources than ch1 and ch2. The underlying typical stellar photosphere is also intrinsically fainter at ch3 and ch4 than at ch1 and ch2, which prevents the ability to detect sources in ch3 and ch4. We finally extracted  4164, 2282, 2018 and 1587 sources in the ch1, ch2, ch3 and ch4 images, respectively. The IRAC data of the four bands were merged, by matching the coordinates with a matching  radius of 1.2 arcsec. We restricted our catalogue to sources having uncertainty $\leq$ 0.2 mag in all the bands in order to ensure good photometric accuracy. Thus, our final IRAC catalogue contains photometry of 707 sources which were detected in all the four IRAC bands.

\citet{snider2009ApJ...700..506S} have reported the detection of 186 sources in an area of $\sim$400 arcmin$^2$ using MIPS and the same IRAC observations. Forty five of them were classified as candidate YSOs. Thirty one sources out of 45 are located in the FoV of the present work. Fifteen out of these 31 sources are classified as candidate YSOs candidates in the present work. The NIR-TCD for the candidate YSOs identified by \citet{snider2009ApJ...700..506S} reveals that a few of them are located below the  locus of the unreddened Classical T-Tauri Stars (CTTS) by \citet[][cf. Section \ref{nir_sec}]{Meyer1997AJ....114..288M}. These sources are considered as non-YSOs in the present study. A comparison of the present photometry with that of \citet{snider2009ApJ...700..506S} shown in Fig. \ref{ph_cmp} indicates a systematic shift of 0.1$-$0.5 mag with a large amount of scattering which increases towards longer wavelength. This discrepancy may be due to the different techniques (aperture photometry by \citet{snider2009ApJ...700..506S} versus present PRF fitting) adopted in these two studies.
 
\subsection{Completeness of the photometric data}
The present photometric catalogue may be  incomplete due to various reasons, e.g., crowding of stars, nebulosity and detection limits, etc. In order to evaluate the completeness limit of the photometric data in $J, H,$ \ks, ch1, ch2, ch3 and ch4 bands, we performed  ``ADDSTAR" experiments of DAOPHOT II. The procedures have been outlined in detail by several authors \citep[see, e.g.,][and references therein]{Pandey2001, Sharma2007, Jose2016}. Briefly, in this method artificial stars of known magnitudes and positions are added randomly into the original frame. Around 15$\%$ of the total stars were added to the original frame in such a way that more number of faint stars were inserted without changing the crowding characteristics significantly \citep[see][]{Sagar1991}. The frames were reduced using the same procedure as used for the original frame. The ratio of the number of stars recovered to those added in each magnitude interval gives the completeness factor (CF) as a function of magnitude.  The CF for different ISPI and IRAC bands is given in Fig. \ref{comp_fac}. The $J, H$, \ks, ch1, ch2, ch3 and ch4 data are found to be 90\% complete at the level of 18.0, 17.7, 17.2, 14.3, 13.7, 13.3 and 11.3  mag, respectively.    

\subsection{Radio Continuum Observations}
 In order to probe the properties of ionized gas, we have carried out 1280 MHz radio continuum interferometric observations on 2007 December 24 (proposal code 13MRS01) using the GMRT array. GMRT consists of thirty antennae, each with  a diameter of 45-m, arranged in a ``Y" shaped configuration. These antennae are distributed in a hybrid configuration with six antennae along each of the three radial arms (with arm length of $\sim$14 km) which provide the longest baseline of $\sim$25 km and twelve antennae randomly distributed in the central region in a compact area (referred to as the central square) of 1 km $\times$ 1 km near the center of ``Y", which provide the shortest baseline of $\sim$100 m. GMRT provides full a FoV of 26\farcm2$\pm$2\arcm at 1280 MHz. Further details about GMRT can be found in \citet{swarup1991}. The radio sources 3C48 and 3C286 were used as the primary flux calibrators, whereas 0739+016 was used as a phase calibrator.

We analysed the data using AIPS software. The data were iteratively cleaned and calibrated by using the tasks ``TVFLG" and ``CALIB-GETJY-SETJY", respectively. After having removed most of the bad data (encompassing bad antennae, bad baselines, RFI affected data and so on), and obtaining a satisfactory calibration, the target source was split from the entire dataset by using the task ``SPLIT". Finally, facet imaging was done by using tasks ``SETFC" and ``IMAGR" on this split target data in order to get the image of the region. Since our object is located near the Galactic plane, we had to take into account the system temperature corrections for GMRT. At meter wavelengths, there is a large amount of noise from the Galactic plane, therefore, effective temperature of the antennae increases, which has to be corrected. This is done by obtaining the sky temperature map of \citet{Haslam1982} at 408 MHz, and by rescaling the image by a scaling factor which is just the ratio of the system temperature toward the target source and the flux calibrator. Details of the procedure have been discussed in \citet{mallick2012}. Fig. \ref{nvss_ir_cc_image} shows the GMRT radio continuum contour map of the S311 region  at 1280 MHz.  In addition to our data at 1280  MHz, we also obtained the NRAO-VLA Sky Survey (NVSS) archive image at 1400 MHz. 

\section{Analysis and results}
\label{ana_res}
\subsection{Identification and classification of YSOs}\label{ysos_iden} 
 Infrared (IR) data are a very useful tool to detect YSOs and study their nature in clusters/SFRs. YSOs exhibit strong IR excess due to the presence of circumstellar disks and envelopes, hence they can be separated out by using the NIR and MIR observations. In this section, the following procedures are adopted to identify and classify YSOs in the region.

\subsubsection{On the basis of $J, H,$ {\ks~} data}\label{nir_sec} 
The NIR-TCD, shown in Fig. \ref{nir_tcd}a, was used to identify the YSO population in the S311 region.  The evolutionary status of the candidate YSOs defines their locations on the NIR-TCD. 
The red-continuous and green-dashed curves represent the unreddened MS and giant branch \citep{Bessell1988PASP..100.1134B}, respectively. The dotted line indicates the locus of the unreddened CTTSs \citep{Meyer1997AJ....114..288M}. All the curves and lines are in the CIT system. The parallel dashed lines are the reddening vectors drawn from the tip (spectral type M4) of the giant branch (left reddening line), from the base (spectral type A0) of the MS branch  (middle reddening line) and from the tip of the intrinsic CTTS line (right reddening line). The extinction ratios $A_J$/\av = 0.282, $A_H$ /\av = 0.175 and $A_K$/\av = 0.112 have been taken from \citet{rieke1985}. The TCD is divided into `F', `T' and `P' regions \citep[cf.][]{ojha2004a,ojha2004b}. The sources located in the `F' region could either be field stars (MS stars, giants) or Class III and Class II sources with small NIR-excesses, whereas the sources distributed in the `T' and `P' regions are considered to be mostly  CTTSs (or Class II objects) with relatively large NIR-excesses and likely Class I objects, respectively (for details see, e.g., \citealp{Pandey2008MNRAS.383.1241P}). However, it is worthwhile mentioning that \citet{robitaille2006} have shown that there is a significant overlap between protostars and CTTSs.  Using the NIR-TCD for the S311 region  (Fig. \ref{nir_tcd}a),  we have detected 71 Class II sources falling in the `T' region which are shown as filled green triangles.} Two \halpha emission stars (see, Section \ref{slitless}) are also marked on the NIR-TCD.

\subsubsection{ On the basis of IRAC ch1, ch2, ch3, ch4 data}\label{ch1234_ysos}
 We used IRAC detections in all ch1, ch2, ch3 and ch4 bands with the uncertainty less than 0.2 mag to identify Class I and Class II sources associated with the region following the schemes described in detail by \citet[][see Appendix A$-$Phase 1]{Gutermuth2009ApJS..184...18G}. Various non-stellar contaminations, which include emission from Poly-cyclic aromatic hydrocarbons (PAH) dominated features, likely star-forming galaxies, and weak-line active galactic nuclei (AGNs) \citep[see][]{Gutermuth2008b}, are eliminated by using these schemes. Out of 707 common sources detected in all IRAC bands, 454 were found to be contaminants. The remaining 253 sources were used to identify Class I and Class II sources based on the [3.6] $-$ [5.8]/[4.5] $-$ [8.0] TCD shown in Fig. \ref{nir_tcd}b, where Class I and Class II sources are shown as filled red squares and filled green triangles, respectively.  Using this scheme we identified a total of 33 candidate YSOs: 4 Class I and 29 Class II sources.

\subsubsection{Additional YSOs on the basis of $H$, \ks, 3.6 and 4.5 \mum data}\label{hk_ch12_ysos}
Due to higher extinction, $J$ band photometry detects fewer sources as compared to $H$ and \ks -bands. Therefore, $H$ and {\ks~} data were cross-matched with a matching radius of 1 arcsec. Thereafter, this $H,$  {\ks~} catalogue  was matched with the clean ch1, ch2 photometric data with a matching radius of 1.2 arcsec, which yields 453 sources in common. 
 The YSOs detection using all four IRAC bands is limited by lower sensitivity and enhanced nebulosity in the ch3 and ch4 pass-bands. In order to detect the missing candidate YSOs we used the $H$, \ks, ch1, and ch2 common source catalogue to further identify additional Class I and Class II sources using the classification scheme developed by \citet[][see Appendix A$-$Phase 2]{Gutermuth2009ApJS..184...18G}. The dereddened (\ks - [3.6])$_0$ and ([3.6] - [4.5])$_0$ colours were calculated by using the extinction laws provided in \citet{Flaherty2007ApJ...663.1069F}. We used the extinction map (cf. Section  \ref{redd_map}) to deredden the individual sources. In order to separate probable faint extra-galactic contaminants, we applied the criteria on the dereddened ch1 magnitude as ch1 $\leq$ 15 mag and ch1 $\leq$ 14.5 mag for Class I and Class II sources, respectively as suggested by \citet{Gutermuth2009ApJS..184...18G}. Fig. \ref{nir_tcd}c shows the location of these Class I and Class II sources, which have been marked as  filled red squares and filled green triangles, respectively on the (\ks $-$ [3.6])$_0$/([3.6] $-$ [4.5])$_0$ TCD. Using this scheme we identified 7 Class I and 61 Class II sources.
 
\subsubsection{Final catalogue of YSOs}
The catalogues of candidate YSOs produced by using the above mentioned three schemes are not mutually exclusive. Therefore, we obtained  a final catalogue of 125 candidate YSOs after merging the above three catalogues. Out of 125 candidate YSOs, 8 sources are found to be Class I and the remaining 117 sources are Class II. It is worth mentioning that our NIR observations are $\sim$2 magnitude deeper than the 2MASS data. A Majority of the candidate YSOs were identified from our deep NIR observations and $H, Ks,$ ch1 and ch2 catalogue. The sample of the final catalogue of identified candidate YSOs is given in Table \ref{yso_cat}.  The full catalogue is available in an electronic form only.

\subsection{Reddening and extinction map of the region}\label{redd_map}
\begin{figure*}
\centering
\includegraphics[width=9cm,angle=0]{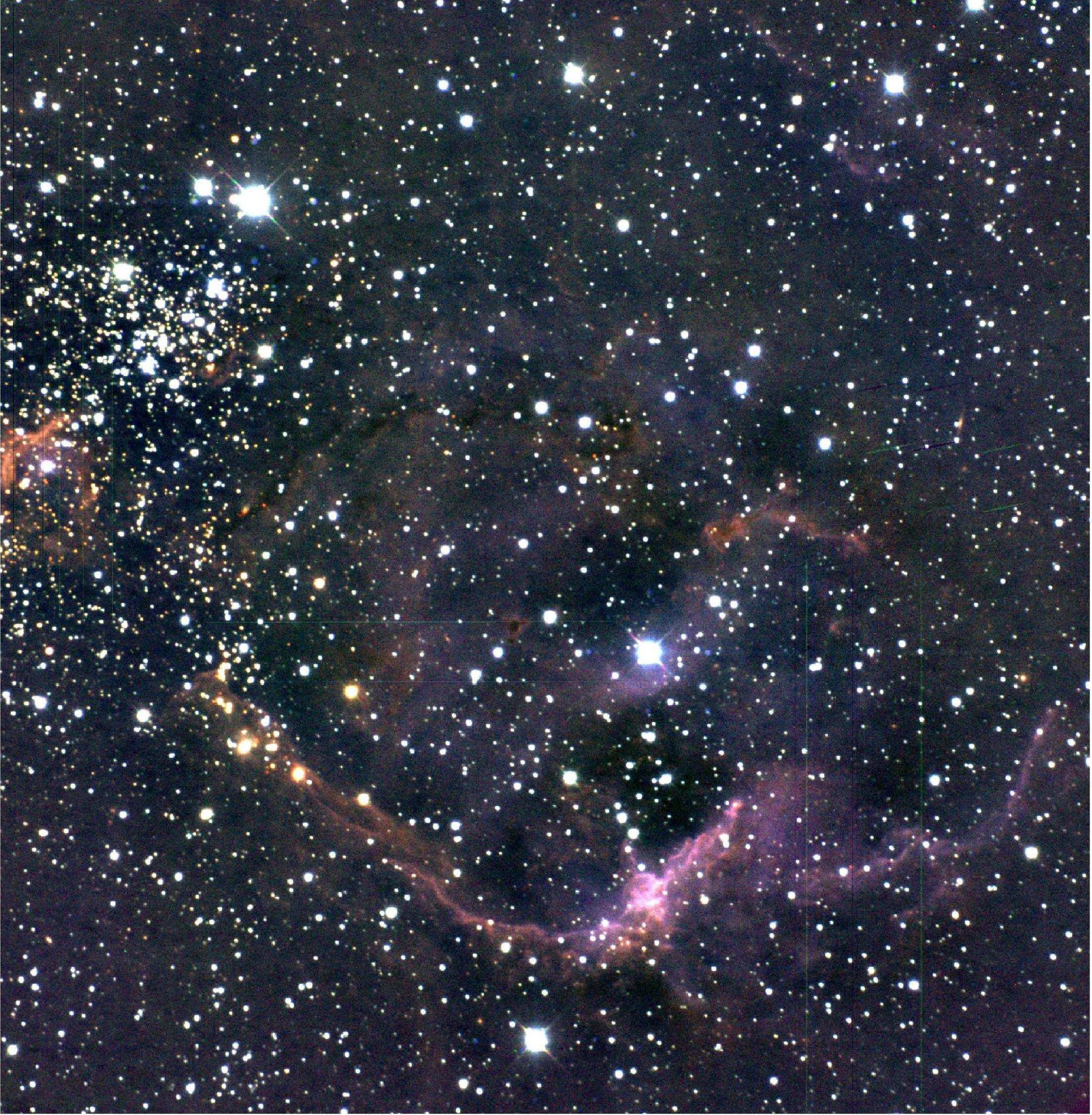}
\caption{\label{nir_cc_im} ISPI NIR colour-composite image of the S311 region made by using the $J$ (blue), $H$ (green) and {\ks~} (red) images. The field of view is $\sim$9.2$\times$9.4 arcmin$^2$. North is up, and east is to the left.}
\end{figure*}

\begin{figure*}
\centering
\includegraphics[width=9cm]{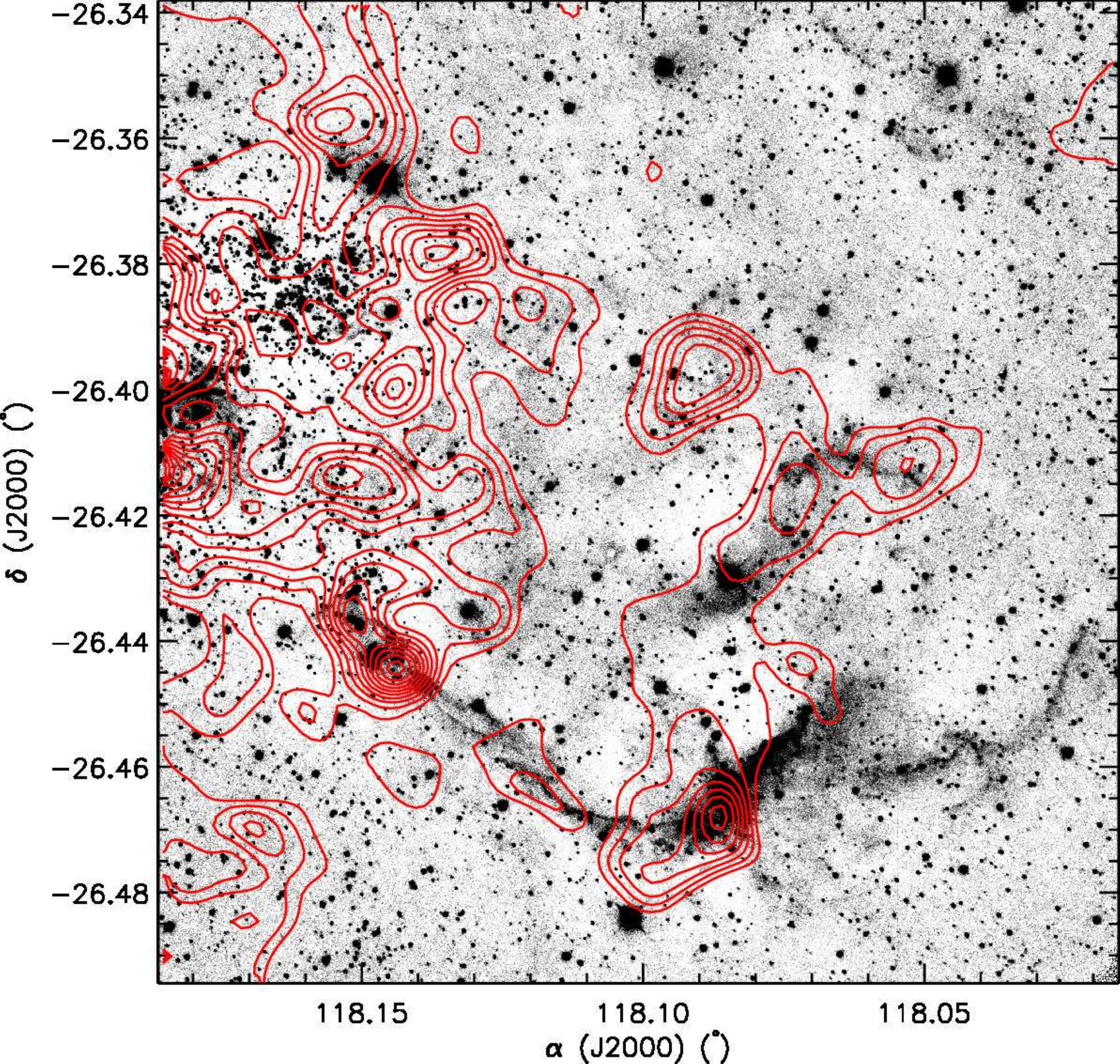}
\caption{\label{ext_map} \ks-band image of the S311 region. Overplotted are the extinction contours generated from the $(H -$ \ks$)$ colours of the ISPI data. The contours are plotted in the range 0.6 mag $\le$$A_K$$\le$ 1.5 mag with an increment of 0.1 mag.}
\end{figure*}

The extinction towards a SFR can arise due to two main reasons: (1) the general interstellar medium in the foreground of the region, (2) the localized cloud associated with the SFR. 
The S311 region, as is clearly evident from Figs. \ref{eso_im}, \ref{nir_cc_im}, and \ref{ext_map}, is still embedded in the parent molecular cloud and can have a variable extinction. The reddening $E(B - V)_{min}$ for the S311 region was estimated in \citet{Yadav2015} by using the $(U-B)$/$(B-V)$ TCD as $\simeq$0.55 mag. The patchy distribution in Figs. \ref{eso_im} and \ref{nir_cc_im} could be due to high extinction caused by the presence of dense molecular material. To characterize the extinction in the S311 region, we derived an $A_K$ extinction map using the $(H$ $-$ \ks$)$ colours of the background stars. To avoid the contribution of NIR-excess sources, we excluded the candidate YSOs identified in Section \ref{ysos_iden}. We generated the extinction map of the region using the ``nearest neighbourhood" method as described in \citet{gutermuth2005} and \citep{Jose2016}. At each position of a grid size of 10 arcsec, we measured the median $(H$ $-$ \ks$)$ value of the five nearest stars. Applying the reddening laws by \citet{Flaherty2007ApJ...663.1069F}, we converted the $(H$ $-$ \ks$)$ colour to $A_K$, using the relation $A_K $ = 1.82 $\times$ $(H$ $-$ \ks$)_{obs}$ $-$ $(H$ $-$ \ks$)_{int}$, where, $(H$ $-$ \ks$)_{int}$ = 0.2 is assumed as the average intrinsic colour of stars in young clusters \citep[see][]{Allen2008ApJ...675..491A,Gutermuth2009ApJS..184...18G,Jose2013}. It is expected that the foreground field stars would affect the extinction map of the S311 region, hence, we excluded the foreground extinction contribution assuming an average foreground reddening towards the region as $E(B-V)$ $\sim$0.55 mag \citep[\av $\sim$1.8 mag, cf.][]{Yadav2015}. To eliminate the foreground contribution while generating the extinction map, we used only those stars which have $A_K$ $>$ 0.2 mag (i.e. \av $>$ 1.8 mag). The final extinction map, having an angular resolution of 10$\times$10 arcsec$^2$ and sensitive down to \av $\sim$16 mag, is shown in Fig. \ref{ext_map}. The central region of S311 seems to be less reddened, as compared to its periphery. 
\subsection{Emission from ionized gas in the S311 region}
\begin{figure*}
\centering
\includegraphics[trim=0.7cm 0.70cm 0.68cm 0.0cm, clip,width=9cm]{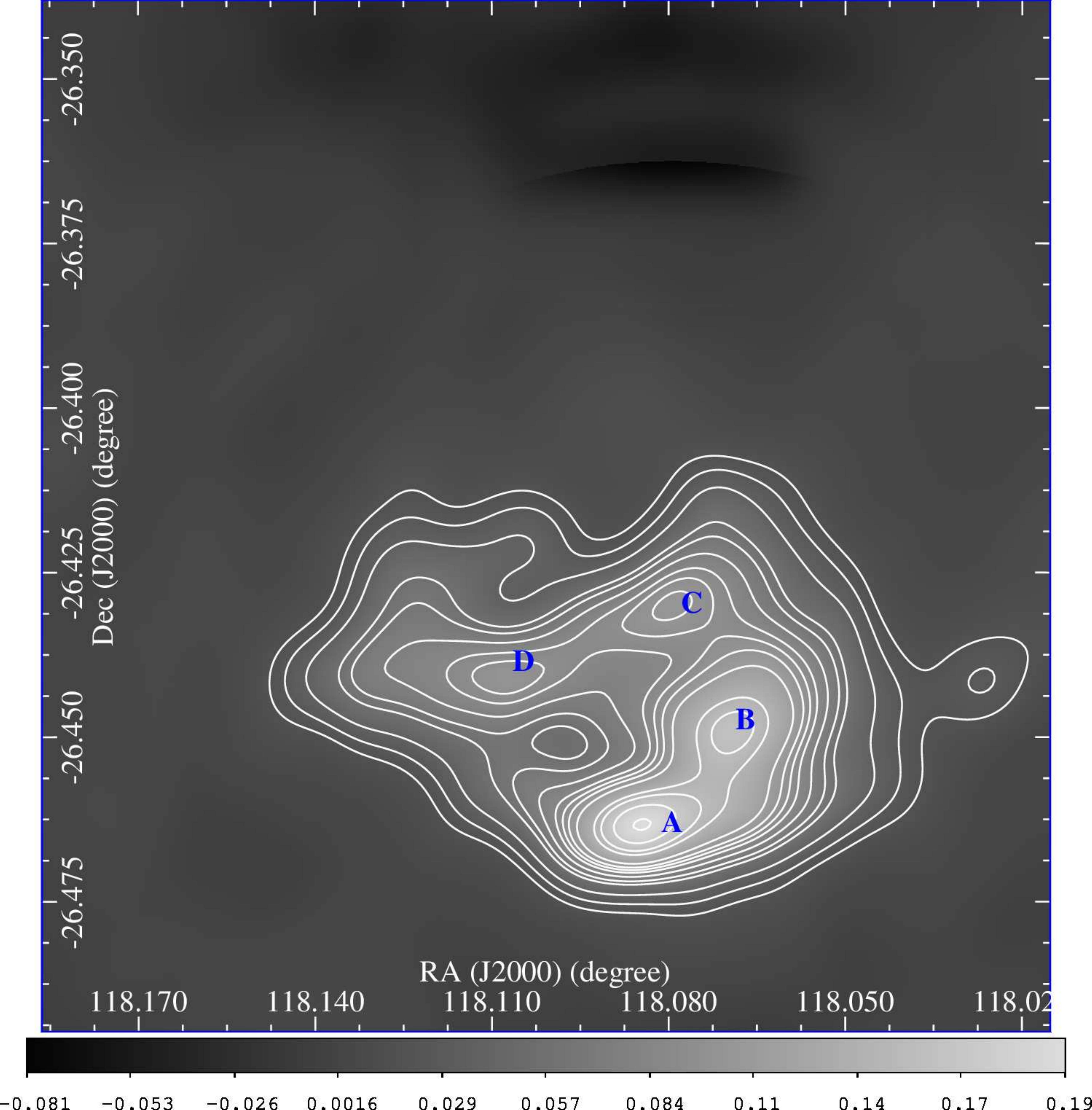} 
\caption{\label{nvss_ir_cc_image}  
GMRT low-resolution contour map of the S311 region at 1280 MHz. The resolution is 43 $\times$ 33 arcsec$^2$ and contours are drawn at 3.8 $\times$ (5, 7, 10, 15, 17, 20, 23, 25, 27, 30, 35, 37, 40, 45, 50) mJy beam$^{-1}$, where $\sim$3.8 mJy beam$^{-1}$ is the rms noise in the map. The labeled axes are in J2000 coordinates. Four clumps/compact \hii regions discussed in the text are labeled as A, B, C and D.}
\end{figure*}

The morphology of the radio continuum emission contours in Fig. \ref{nvss_ir_cc_image} reveals that the southwestern (SW) region is an ionization-bounded zone, whereas the decreasing intensity distribution towards the northeastern (NE) direction indicates that this region could be density bounded.  The integrated flux density from the radio continuum contour map for the \hii region S311 above 5$\sigma$ level is estimated to be 2.71 Jy. Assuming the ionized region to be spherically symmetric and neglecting absorption of ultraviolet radiation by dust inside the \hii region, the above flux density together with an assumed distance of 5.0 kpc, allows us to estimate the number of Lyman continuum photons ($N_{\rm Lyc}$) emitted per second, and hence the spectral type of the exciting star(s). Using the relation given by \citet{Kurtz1994ApJS...91..659K} for an electron temperature of 10500 K \citep{Balser2011ApJ...738...27B}, we estimated $N_{\rm Lyc}$ = 5.1 $\times$ 10$^{48}$, which corresponds to a Zero Age Main Sequence (ZAMS)/MS spectral type of $\sim$O7 \citep{Panagia1973AJ.....78..929P}. On the basis of optical spectroscopy,  \citet{Yadav2015, Walborn1982} and \citet{Crampton1971} estimated the spectral type of the ionizing source of S311 as O6V$\pm$1, which is in fair agreement with the spectral type estimated from the integrated radio continuum flux. Major uncertainty in the above calculation may arise from neglecting the absorption of Lyman continuum photons by the dust. Another plausible reason could be the high resolution of GMRT data which might have filtered out the extended emission. A low resolution survey of the region  by \citet{Vollmer2010A&A...511A..53V} at 4800 MHz  gives an integrated flux of 4.3 Jy, which yields $N_{\rm Lyc} \sim10^{49}$, suggesting an O6 V spectral type for the ionizing source, which agrees well with the spectral class estimated by previous studies. The GMRT radio data reveals four peaks which may be clumps/compact \hii regions, marked as A, B, C, and D in  Fig. \ref{nvss_ir_cc_image}. The physical parameters of these clumps/compact \hii regions were calculated by using ``JMFIT" task in AIPS software and are given in Table \ref{rad_par}. These compact \hii regions might have formed due to the expansion of ionized gas and could be the sites of high-mass stars in their earliest stages.

\subsection{\label{phy_pro} Physical Properties of YSOs}

\subsubsection{Spectral Energy Distribution of YSOs}\label{sed_dis}
\begin{figure*}
\centering
\hbox{
\includegraphics[width=8.0cm]{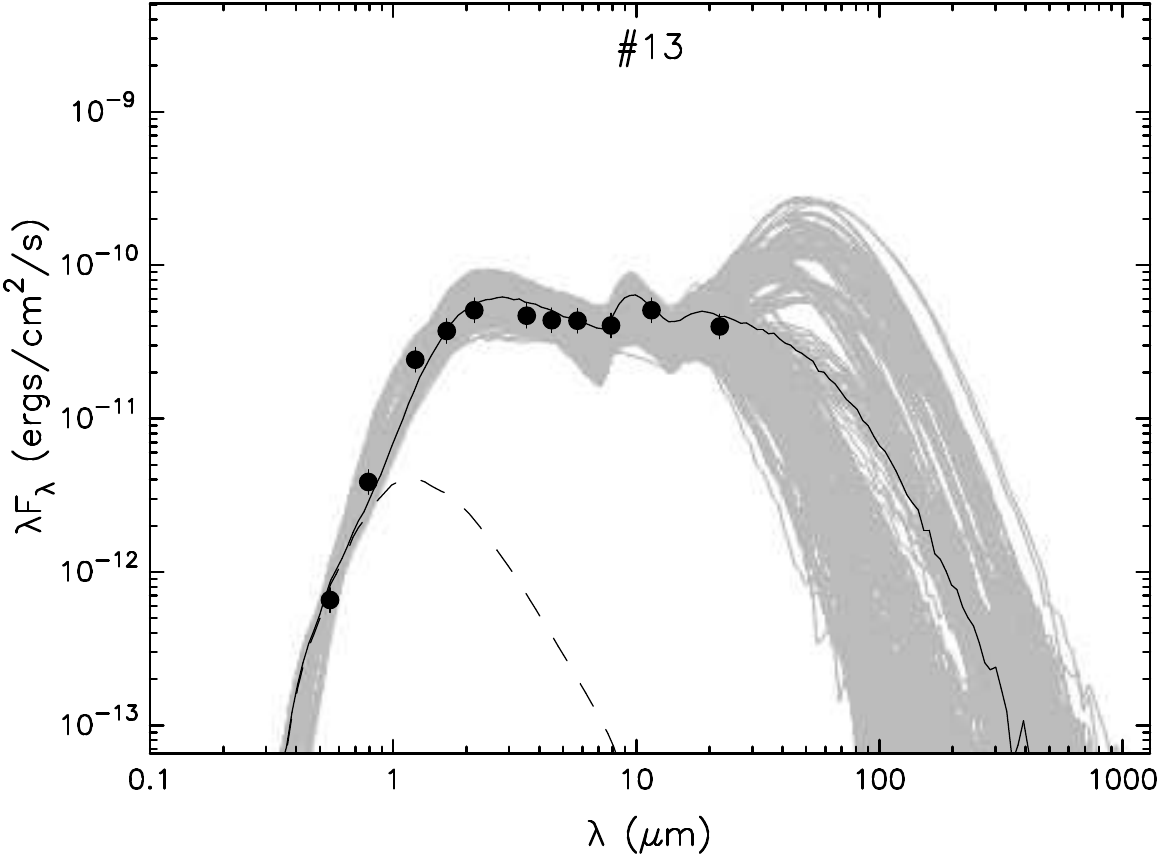}
\includegraphics[width=8.0cm]{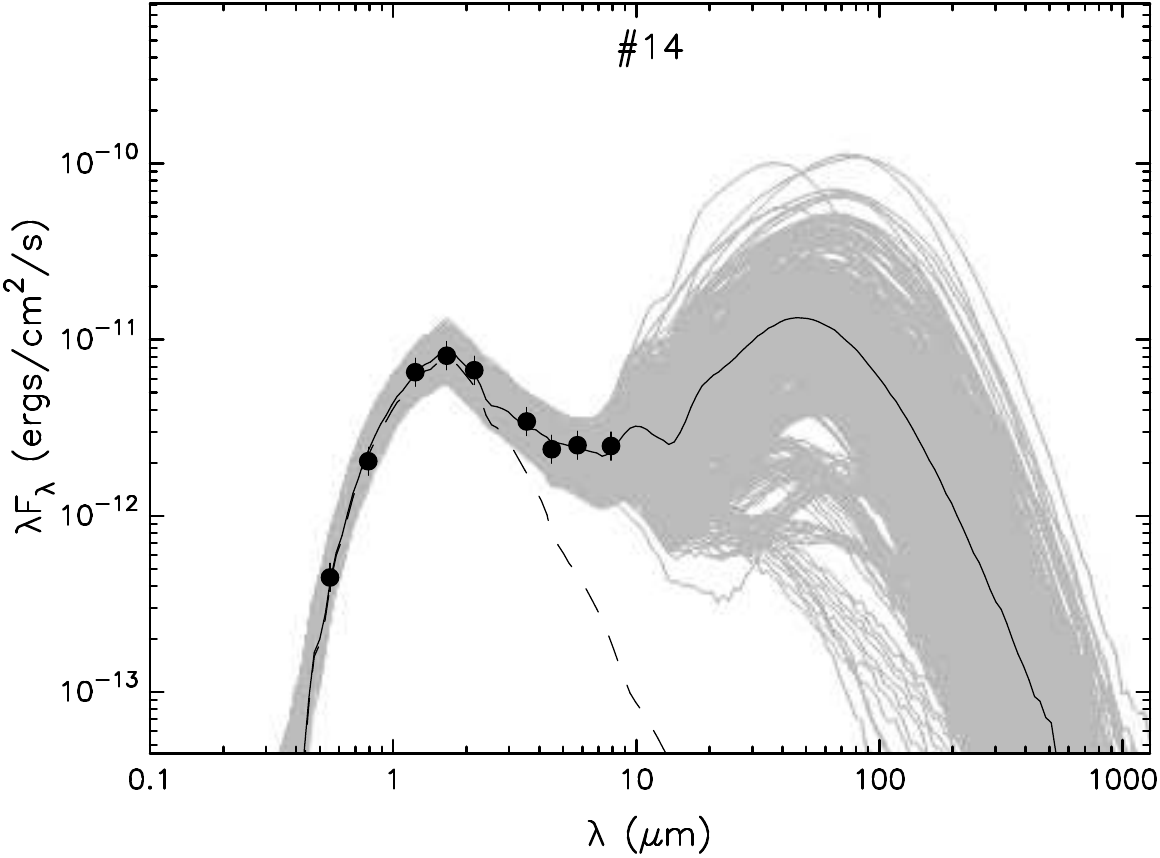}
}
\caption{Sample SEDs for a Class I (left) and Class II (right) candidate YSO by using the SED fitting tools of \citet{robitaille2007}. The solid black line shows the best fit and the gray lines show the subsequent fits. The dashed line shows the stellar photosphere corresponding to the central source of the best-fitting model. The filled circles denote the input flux values.}\label{ysos_sed}
\end{figure*}

 In order to get an idea about the nature of the candidate YSOs identified in the present work, we have carried out SED modeling using the models and fitting tools of \citet{robitaille2006, robitaille2007}. The model grid consists of 20,000 SED models from \citet{robitaille2006} computed by using two-dimensional radiative transfer Monte-Carlo simulations. These models assume an accretion scenario with a central source associated with a rotationally flattened in-falling envelope, bipolar cavities and a flared accretion disk, all under radiative equilibrium. Each YSO model provides the output SEDs for 10 inclination angles, covering a range of stellar masses from 0.1 to 50 \msun. We only select those models that satisfy the criterion: $\chi$$^2$ - $\chi^2_{best}$ $<$ 3$N_{data}$,  where $\chi^2_{best}$  is the goodness-of-fit parameter for the best-fit model and $N_{data}$ is the number of input observational  data points. To constrain the parameters of the stellar photosphere and circumstellar environment, we fit the SEDs to only those sources for which we have a minimum of eight data points in the wavelength range from 0.36 to 8 \mum. Apart from fluxes/magnitudes at different wavelengths as input parameters, two other important input parameters were the range of distance (taken to be 4.5 $-$ 5.5 kpc) and the visual extinction \av (taken to be 1.7 $-$ 16 mag). The upper limit of \av  was considered as 16 mag which is the maximum value of extinction estimated on the basis of the extinction map analysis as discussed in Section \ref{redd_map}.  
 We searched for optical counterparts in $VI$ bands to our sample of 125 candidate YSOs in \citet{Yadav2015}, using a matching radius of 1 arcsec, and found a match for 56 sources. To constrain the parameters of the stellar photosphere and circumstellar environment, we fit the SEDs to only those sources for which we have a minimum of eight data points in the wavelength range from 0.36 to 8 \mum. This yields 16 sources detected in eight or more bands ranging from optical ($UBVI$) to IRAC ch4. The parameters are obtained from the weighted mean and the standard deviation  of these models, weighted by $e^{(-\chi^2/2)}$ of each model, and are shown in Table \ref{phy_par}  which shows that the candidate YSOs have ages and masses in the range of $\sim$0.05$-$5 Myr and $\sim$2$-$6 \msun, respectively. The visual extinction in the direction of these sources varies between 1.8 to 8 mag. Fig. \ref{ysos_sed} shows the SEDs for a sample of Class I (left) and Class II (right) YSO candidates. It’s important to note that the SED fitting only provides an age and mass range for 16 of the 125 source sample, and so the fitted parameters may not be completely representative of the entire sample. Therefore, the CMD in Fig. \ref{op_cmd_ysos} is needed to provide an age an mass range for a larger fraction of the YSO sample (see, Section \ref{cmd_ana}).

\subsubsection{CMDs of YSOs: age and  mass estimation}\label{cmd_ana}
\begin{figure*}
\includegraphics[width=8cm]{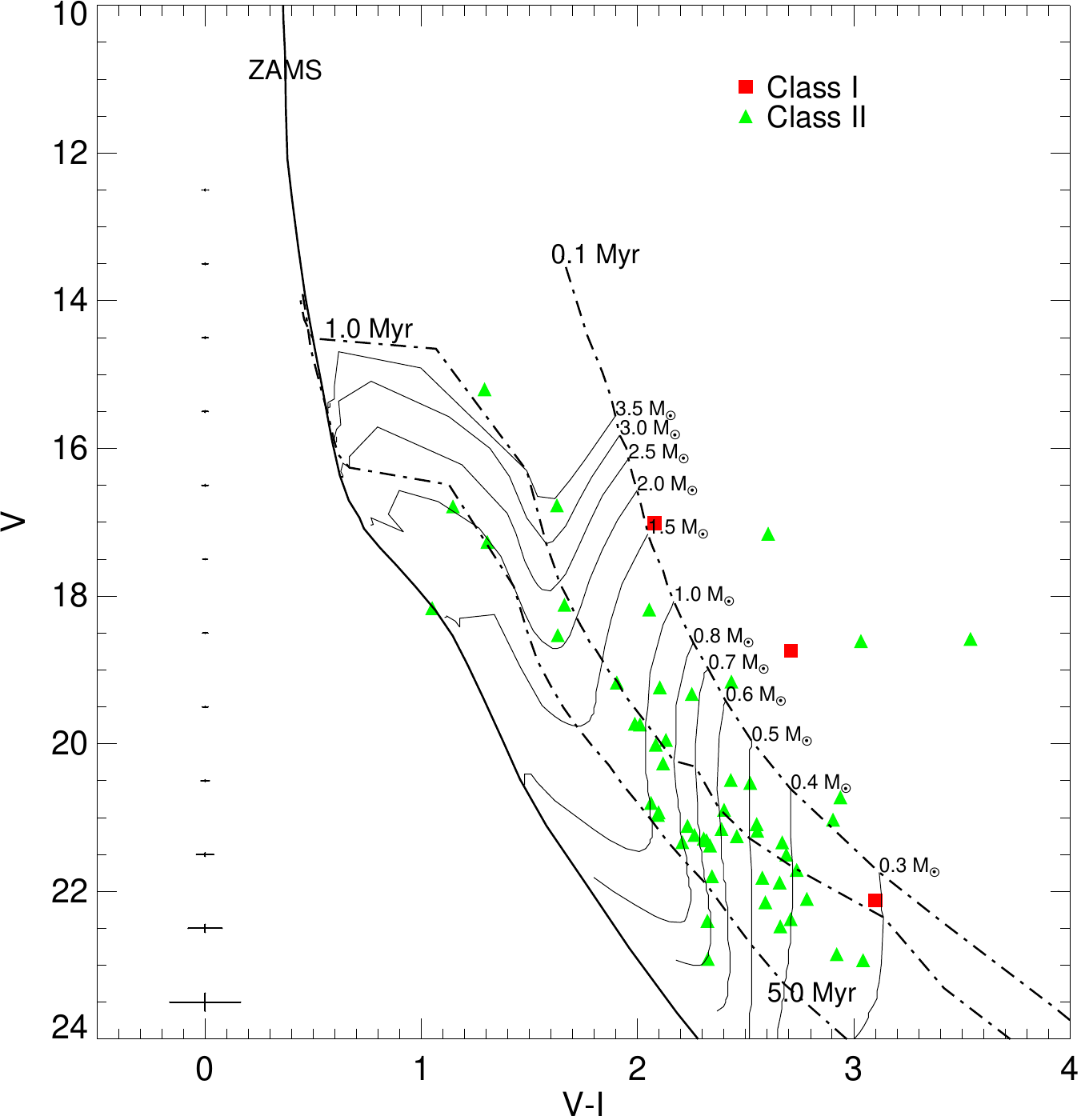}
\caption{\label{op_cmd_ysos} $V$/\vi CMD for the identified candidate YSOs. The PMS isochrones for 0.1, 1, and 5 Myr by \citet{Siess2000A&A...358..593S} and the isochrone for 1 Myr by \citet{Marigo2008A&A...482..883M} are drawn as dotted-dashed and continuous curves, respectively. All the isochrones are corrected for the distance and reddening of S311. The average errors in $V$ and \vi colours are shown on the left side of the figure. Evolutionary tracks for various masses are also shown as thin solid curves.}
\end{figure*}

Using the deep $V$ and $I$ photometry of the 56 identified candidate YSOs from \citet{Yadav2015}, their distribution on the $V/(V - I)$ CMD  is shown in Fig. \ref{op_cmd_ysos}. In Fig. \ref{op_cmd_ysos}, the 1 Myr post-main sequence isochrone (the thick solid curve) by \citet{Marigo2008A&A...482..883M}, which is practically identical to the ZAMS, and pre-main-sequence (PMS) isochrones by \citet{Siess2000A&A...358..593S} for age 0.1, 1 and 5 Myr are also shown. The PMS evolutionary tracks are shown with thin solid curves by \citet{Siess2000A&A...358..593S} for various mass bins. The value of the mass for each track is given towards its right. The isochrones and evolutionary tracks are corrected for the distance of 5.0 kpc and the minimum reddening, i.e., $E(B-V)$ = 0.55 mag. The distribution of candidate YSOs on the $V$/\vi CMD can be used  to estimate their approximate ages and masses. It is evident from Fig. \ref{op_cmd_ysos} that the candidate YSOs in the S311 region are distributed between the PMS isochrones of age 0.1 and 5 Myr and in the mass range of $\sim$0.3$-$3.5 \msun. These are comparable with the lifetime and masses of T-Tauri-stars (TTSs). The age spread indicates a non-coeval star formation in this region. The mean age of the candidate YSOs, excluding 3 stars lying near the MS is found to be 1.53 $\pm$ 1.40 Myr. A sample table giving  the age and mass of candidate YSOs is given in Table \ref{age_mas}. The full table is available in an electronic form only. The ages and masses of the candidate YSOs estimated in Section \ref{sed_dis} are compared with those by the $V$/\vi CMD analysis. The ages from two estimates seem to be correlated with each other, however, the mass estimates by the SED analysis seems to be larger than those by the CMD analysis. 

It is worthwhile to point out that the estimation of the ages and masses of the PMS stars by comparing their locations  in the CMDs with the theoretical isochrones and evolutionary tracks is prone to random as well as systematic errors  \citep{Hillenbrand2005astro.ph.11083H,Hillenbrand2008ASPC..384..200H,Chauhan2009MNRAS.396..964C,Chauhan2011MNRAS.415.1202C}. The systematic errors could be due to the use of different PMS evolutionary models and the error in distance estimation etc. Apart from the latter we presume that our age estimation is affected only by the random errors, since we are using the evolutionary models by \citet{Siess2000A&A...358..593S} to estimate the ages of the candidate YSOs in the region.

The photometric errors, differential reddening, binary population etc. can cause a spread in the CMD. The average photometric uncertainties in $V$ magnitudes and \vi colours estimated by using the ``ADDSTAR'' routine are shown on the left-hand side of the figure. The NIR-TCDs (cf. Fig. \ref{nir_tcd})  of the candidate PMS sources suggest that they are significantly reddened  
and that the region is significantly affected by differential reddening. We note here that the differential reddening does not affect much in the age estimation, since the PMS isochrones are more or less parallel to the reddening vector, however it does in mass estimation. The effect of random errors due to photometric errors and reddening estimation in determination of ages and masses has been estimated by propagating the random errors to the estimates in question by assuming a normal error distribution and by using the Monte-Carlo simulations \citep[cf.][]{Chauhan2009MNRAS.396..964C}. 

The presence of binaries may also introduce errors in the age determination. Binarity will brighten the star, consequently the CMD will yield a lower age estimate. In the case of an equal mass binary we expect an error of $\sim$50 - 60 \% in the PMS age estimation. However, it is difficult to estimate the influence of binaries/variables on the mean age estimation as the fraction of binaries/variables is not known. 

The ages of the young clusters/\hii regions can also be derived from the post-main-sequence age of their most massive members. The ionizing source of S311  is HD 64315, which is an O6-O7V star and seems to be on the MS. 
Hence, the age of the \hii region S311 should be younger or of the order of the MS lifetime of the O6-O7V  star i.e., $\sim$5 Myr \citep{Meynet1994A&AS..103...97M}. Thus, we can put an upper limit to the age of S311 as $\sim$5 Myr.

\subsection{\label{spat_dis} Spatial distribution of YSOs}
\begin{figure*}
\centering
\vbox{
\hbox{
\includegraphics[width=6.0cm]{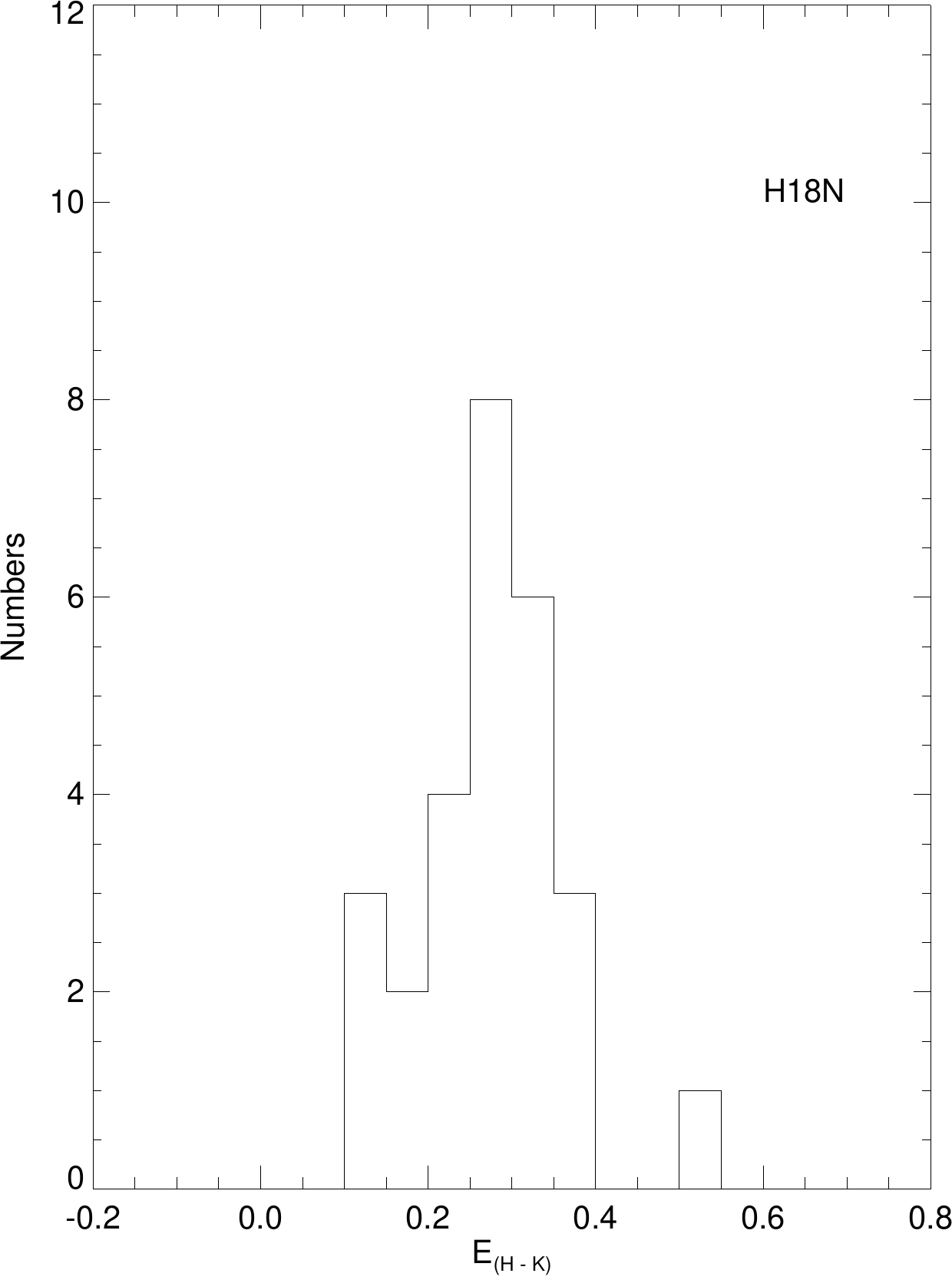}
\includegraphics[width=6.0cm]{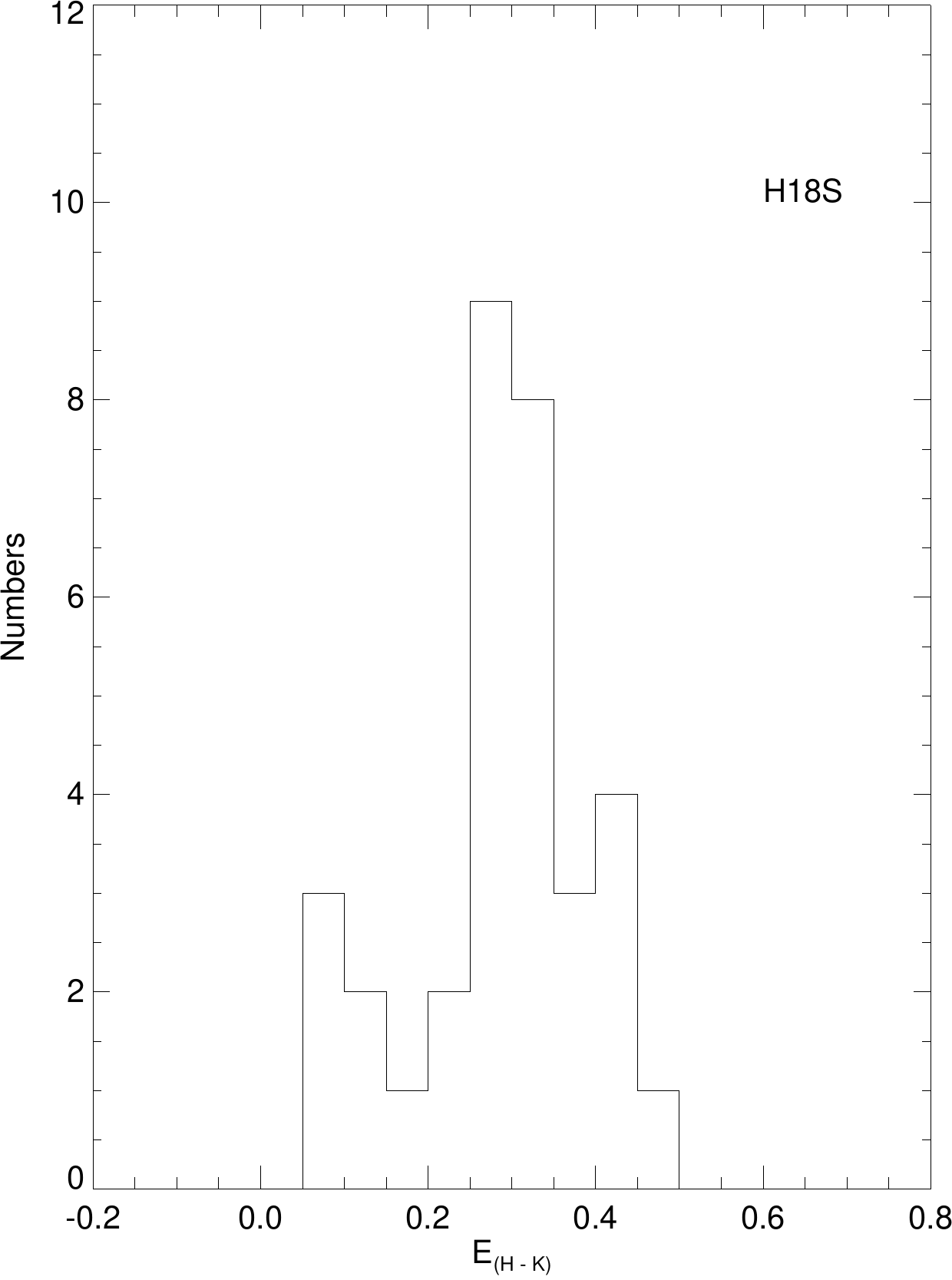}
\includegraphics[width=6.0cm]{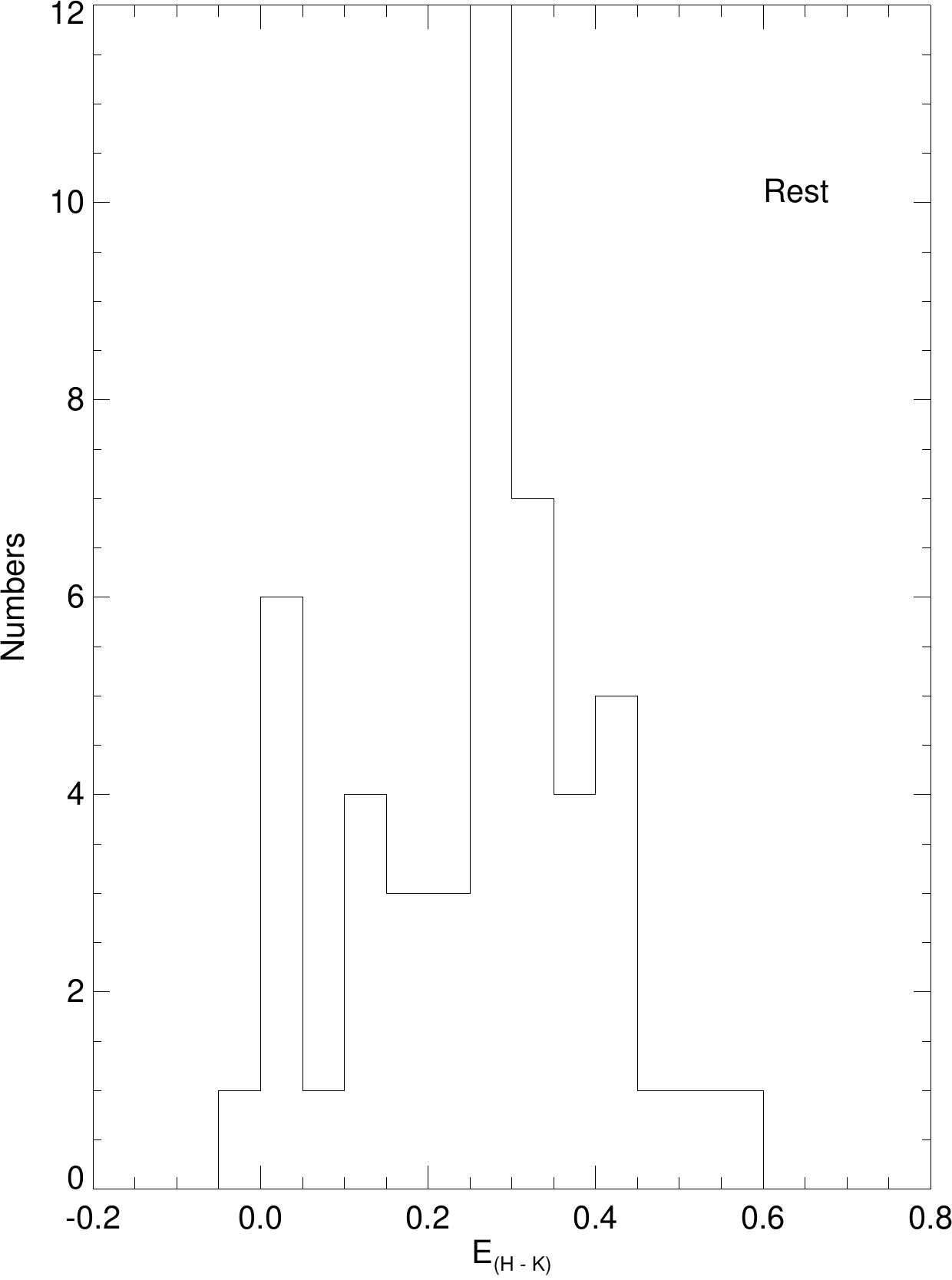}
}
\hbox{
\includegraphics[width=9cm]{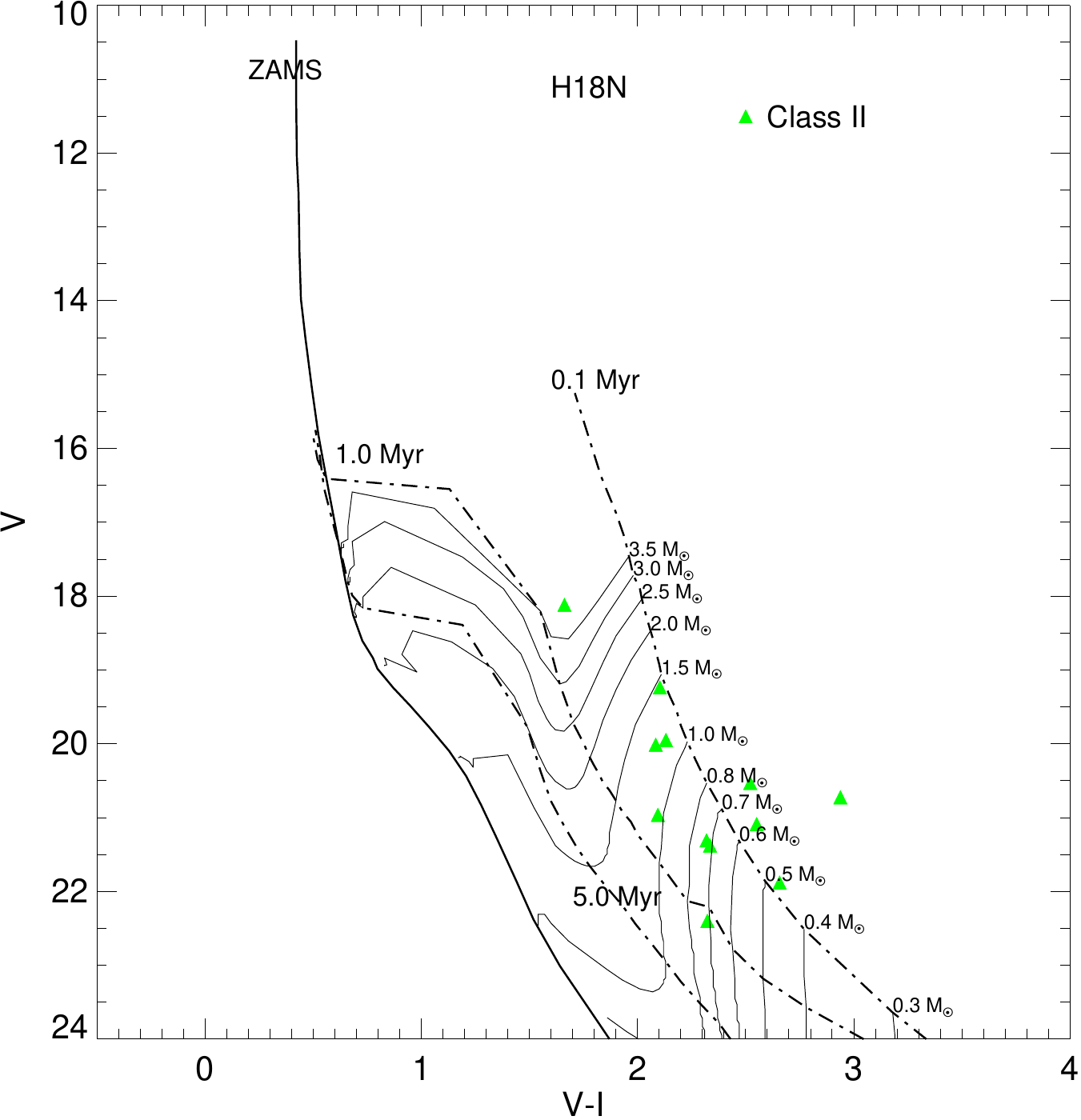}
\includegraphics[width=9cm]{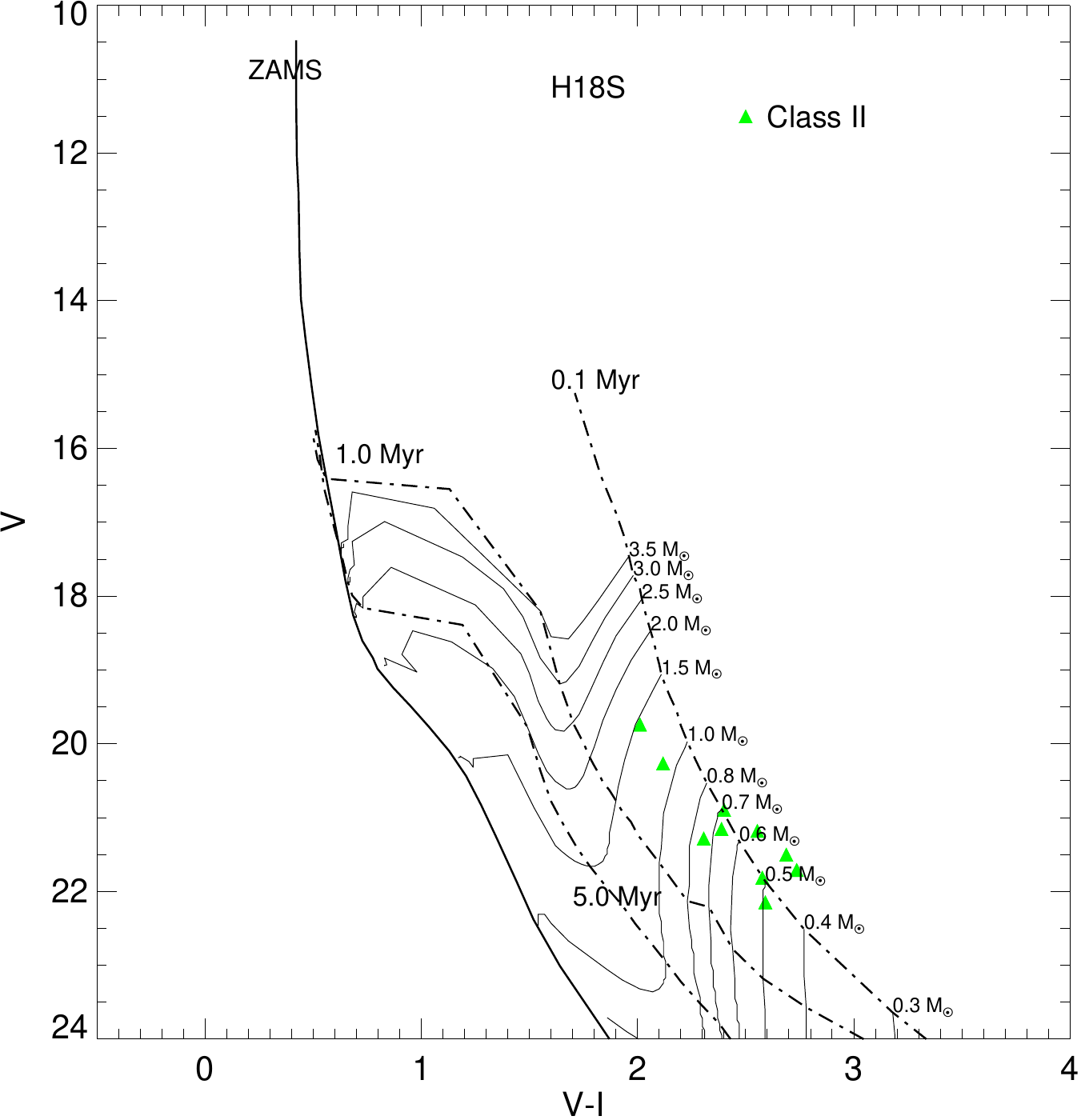}
}}
\caption{Upper panel: $\Delta{(H-K_s)}$ histograms for the sources lying in H18N, H18S and the rest of the area. Lower panel: $V$/\vi CMD for the candidate YSOs located towards the regions H18N and H18S.}\label{ehk_hist}
\end{figure*}

The spatial distribution of candidate YSOs is an important tool to understand the star formation scenario in the region. Fig. \ref{spatial_dist} shows the spatial distribution of the candidate YSOs  on the $K_s$$-$band image along with the 8 $\mu$m contours taken from {\it Spitzer} IRAC ch4 and the radio contours at 1280 MHz. The 8 $\mu$m contours reveal a rather symmetric distribution around the ionizing source HD 64315. It is interesting to note that almost all the Class I sources are distributed along the  8 $\mu$m contours, whereas a significant number of Class II candidate YSOs are clustered towards the eastern  border of the symmetrical distribution of the 8 $\mu$m emission. A significant number of the candidate YSOs are located towards the H18 region. To check whether all of them belong to the same population, we divided the YSOs sample into three sub-samples, namely, H18N, H18S and the remaining sample of the candidate YSOs. The H18N and H18S sub-regions are marked in Fig. \ref{spatial_dist}. We estimated the NIR-excess $\Delta$$(H - K_s)$, which is defined as the horizontal distance of the sources from the left-most reddening vector in the NIR-TCD (see Fig. \ref{nir_tcd}a). The $\Delta$$(H - K_s)$ distribution for the three sub-regions is shown in the upper panel of Fig. \ref{ehk_hist}, which shows a similar distribution for all the three regions. Since $\Delta$$(H - K_s)$ is an indicator of age \citep{Hernandez2007}, we can safely presume that the age distribution of the candidate YSOs in the three sub-regions is identical. In the lower panel of Fig \ref{ehk_hist}, we have also plotted a $V/(V-I)$ CMD for the candidate YSOs lying towards the H18N and H18S regions and isochrones corrected for the distance and reddening of H18 (i.e., distance=11 kpc and $E(B-V)$=0.60 mag), which reveals that all the candidate YSOs become younger than 1 Myr. This seems very much unlikely for Class II objects. Hence, we presume that all the candidate YSOs are located at the distance of S311, i.e., at the distance of 5 kpc and belong to the same population. 

\begin{figure*}
\centering
\hbox{
\includegraphics[width=6.3cm]{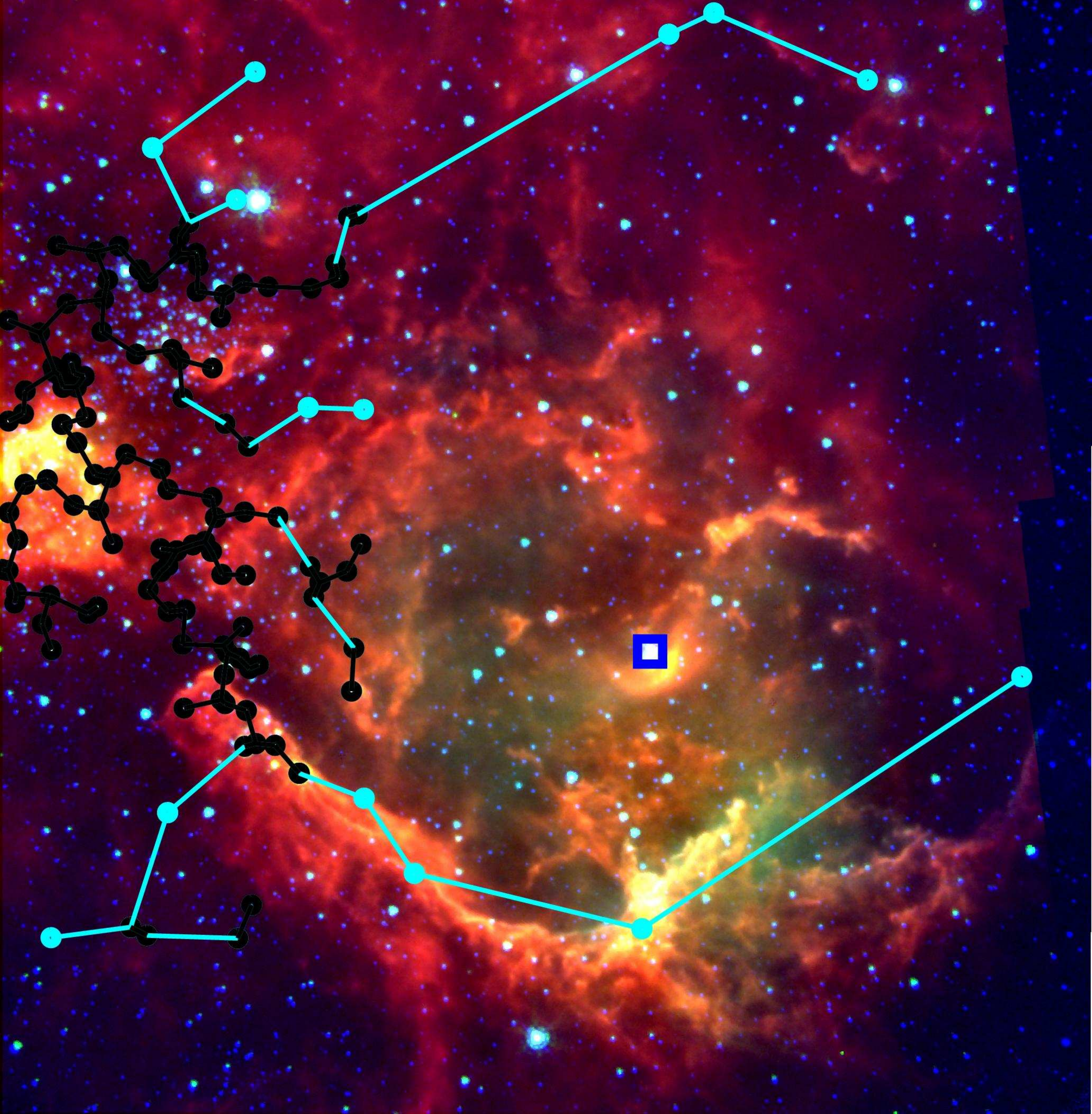}
\includegraphics[width=6.0cm]{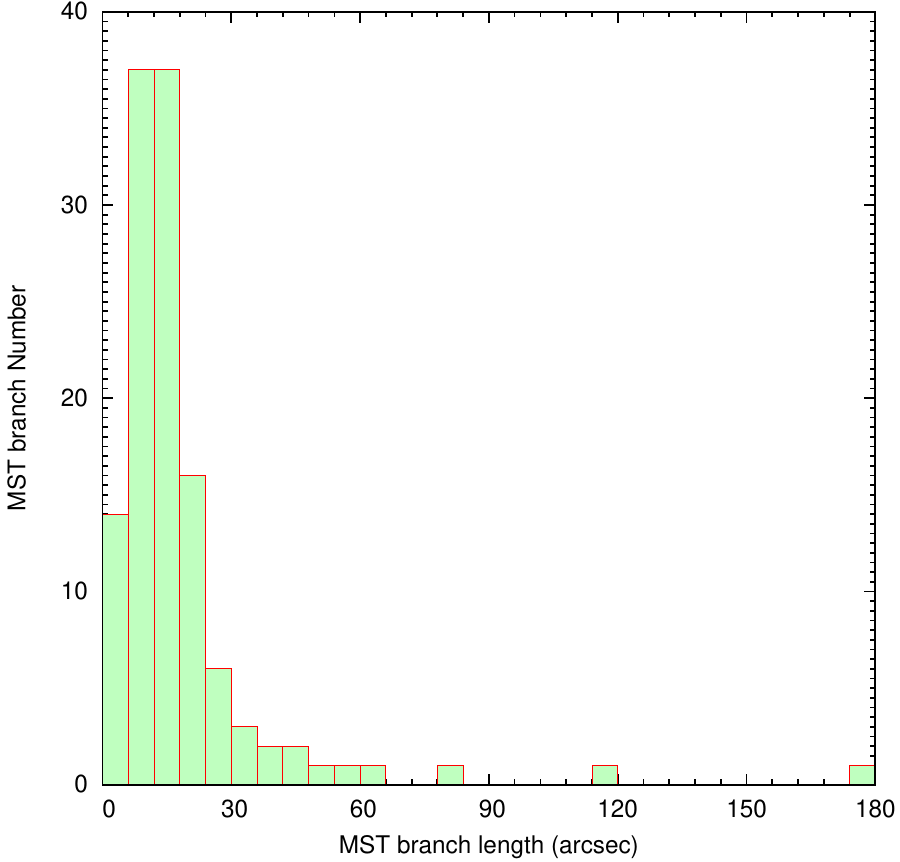}
\includegraphics[width=6.0cm]{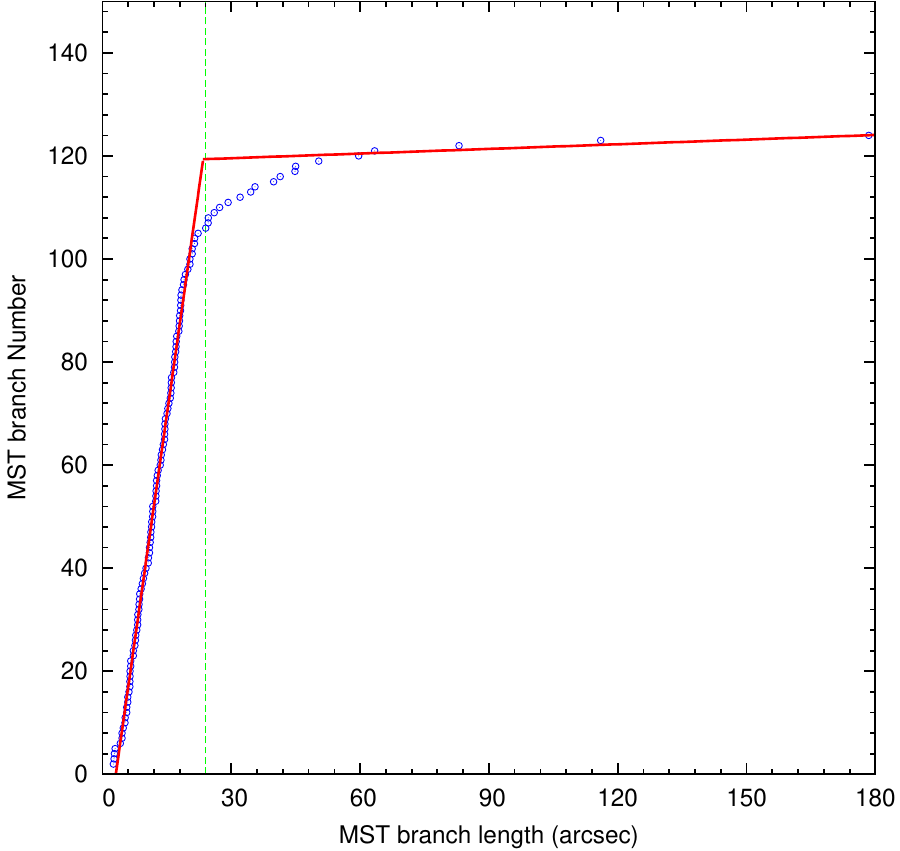}
}
\caption{\label{mst_cdf} Left panel: MST obtained with all the candidate YSOs, whose positions are marked by filled circles. The background image is a colour-composite image made by using {\ks~} (blue), ch1 (green) and ch4 (red) band images. The branches smaller than the critical length ($\sim$0.7 pc) and the connecting points are marked by black continuous lines and black dots, respectively; those larger than the critical length are shown by cyan points connected with cyan continuous lines. Middle panel: Histogram of the MST branch length. Right panel: cumulative distribution of the MST branch length. The straight lines represent the linear fit to the points smaller and larger than the chosen critical branch length.}
\end{figure*}

 The candidate YSOs in Fig. \ref{spatial_dist} are distributed in a clumpy manner. To isolate these cores/clumps of candidate YSOs from the non-uniform background due to the clumpy nature of the molecular cloud, we followed the statistical approach by generating a minimal spanning tree (MST) for the candidate YSOs, as already has been discussed in detail by \citet{gutermuth2005,Gutermuth2008b,Gutermuth2009ApJS..184...18G}. Using the identified candidate YSOs we estimated the projected separation from each YSO to its neighbours. Fig. \ref{mst_cdf} (left panel) shows the MST of the YSO positions. Black points and black MST connections are more closely spaced objects than the critical length as measured in the cumulative distribution of MST branch lengths (right panel of Fig. \ref{mst_cdf}). The critical length is estimated as described by \citet{Gutermuth2009ApJS..184...18G} by using the histogram method (Fig. \ref{mst_cdf}; middle panel) and the cumulative distribution (Fig. \ref{mst_cdf}; right panel) which comes out to be $\sim$0.7 pc. The candidate YSOs and MST branches shown with cyan,  
indicate isolated objects having distances more than the critical length.  
We can easily see the close grouping of candidate YSOs on the eastern side of the \hii region.

\subsection{Mass function of identified YSOs population}
The initial distribution of stellar masses in a SFR is known as the initial mass function (IMF) and together with the star formation rate and stellar density, it dictates the evolution and fate of star clusters as well as of galaxies \citep{Kroupa2002Sci...295...82K}. 
The mass function (MF) is often expressed by the power law, $N$(log $m$) $\propto$ m$^{\Gamma}$, and its slope is given as $\Gamma$ = d log $N$(log $m$)/d log $m$,  where $N$(log $m$) is the number of stars per unit logarithmic mass interval. For the mass range 0.4 $<$ $M/$\msun $<$ 10, the classical value derived by \citet{Salpeter1955ApJ...121..161S} for the slope of MF is = -1.35. 

\begin{figure*}
\centering
\includegraphics[trim=1.8cm 2.7cm 0cm 0cm, clip, width=9cm]{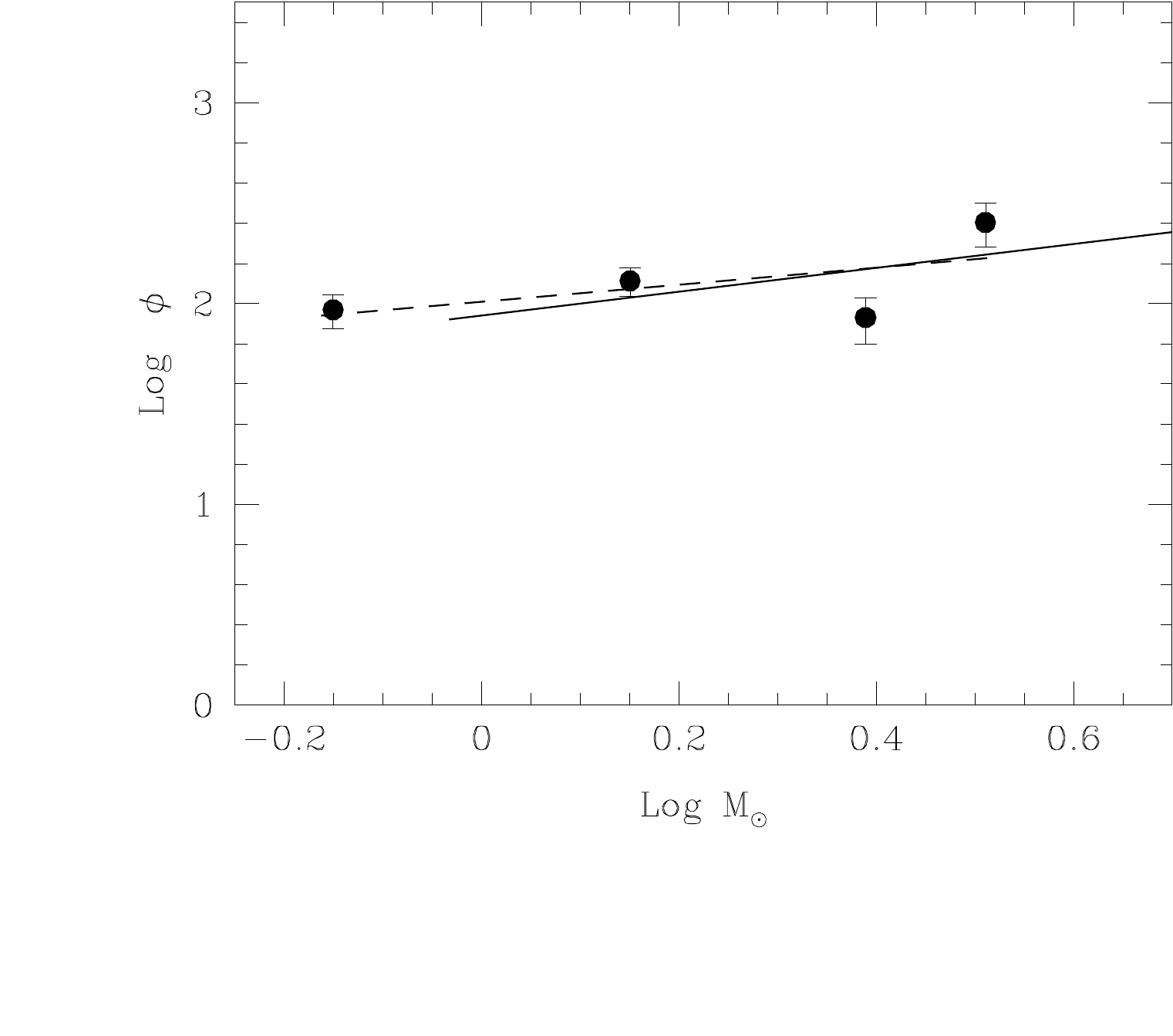} 
\caption{The IMF of the candidate YSOs in the S311 region. The $\phi$ represents $\frac{N}{\rm d log m}$. The least-square fit to the distribution in the mass range 0.3$-$3.5 M$_{\odot}$ is shown by a dashed line yields a slope of 0.42 $\pm$ 0.43. The mass function slope in the mass range 1.0$-$3.5 \msun, shown by continuous line, is found to be 0.59 $\pm$ 1.16.}\label{ysos_imf} 
\end{figure*}

Since our NIR data are deeper, we expect to have a better detection of candidate YSOs towards the fainter end in comparison to the optical data. Therefore, we estimated the IMF of the identified YSOs using the NIR data. Fig. \ref{nir_cmd_ysos} shows the $J/(J-H)$ CMD for the candidate YSOs identified in Section \ref{ysos_iden}. The average age of the candidate YSOs is $\sim$1.53 Myr (Section \ref{phy_pro}). The thin and thick curves represent the 1 Myr PMS isochrone by \citet{Siess2000A&A...358..593S}  and \citet{Baraffe1998A&A...337..403B,Baraffe2003A&A...402..701B}. Both the isochrones are corrected for the cluster distance and minimum reddening. The masses of the probable YSO candidates can be estimated by comparing their locations on the CMD with the evolutionary models of PMS stars. To estimate the stellar masses, the $J$ luminosity is recommended rather than that of $H$ or $K_s$, as $J$ is less affected by the emission from circumstellar material \citep[][]{Bertout1988ApJ...330..350B}. The dashed oblique lines denote the reddening vectors for PMS stars of 1 Myr having masses in the range of 0.3 $-$ 3.5 \msun. Fig. \ref{nir_cmd_ysos} reveals that the majority of the Class II candidate YSOs have masses in the range of 0.5 $-$ 3.5 \msun. Fig. \ref{ysos_imf} shows the derived mass function for the candidate YSOs in the mass range 0.5 $\le$ $M/$\msun $\le$ 3.5, which is found to be rather flat ($\Gamma=0.42\pm0.43$). 
Since our data are complete up to $\sim$90$\%$ for $J$ $\sim$18.0 mag, the present MF estimation  
considering the extinction in the region, should not be affected by the incompleteness down to 1.0 \msun. The MF in the mass range 1.0 $\le$ $M/$\msun $\le$ 3.5 comes out to be 0.59 $\pm$ 1.16. Here we would like to point out that the present MF is only for Class II sources. Several Class III as well as Class II candidate YSOs having very small NIR excess are not included in the sample. We have estimated the total number of probable candidate YSOs in the mass range of 1 $\le$ M/\msun $\le$ 3.5 using the Salpeter MF slope and a constant `A' as described by \citet{snider2009ApJ...700..506S}. However, the S311 region has only one O type star, so we scaled down the constant value for one O type star. The constant `A' derived by \citet{snider2009ApJ...700..506S} was for two O type stars in the region. The estimated number of candidate YSOs in the mass range of 1.0 $\le$ M/\msun $\le$ 3.5 is found to be 206. In Section \ref{ysos_iden} we have identified 125 candidate YSOs, out of which 71 have mass range of 1.0 $\le$ M/\msun $\le$ 3.5. Thus the total number of estimated candidate YSOs yield a fraction of $\sim$35 \% Class II candidate YSOs in the mass range of 1.0 $\le$ M/\msun $\le$ 3.5. The fraction (35 \%) of candidate YSOs suggest an age of $\sim$4 Myr for the S311 region \citep[cf.][]{Hernandez2008}, which is comparable with the estimated age of the region and the age of the ionizing source in the region. 

\section{Discussions}
\label{discu}
\begin{figure*}
\centering
\hbox{
\includegraphics[trim=0.0cm 0.3cm 0.0cm 0.0cm, clip, width=8.4cm]{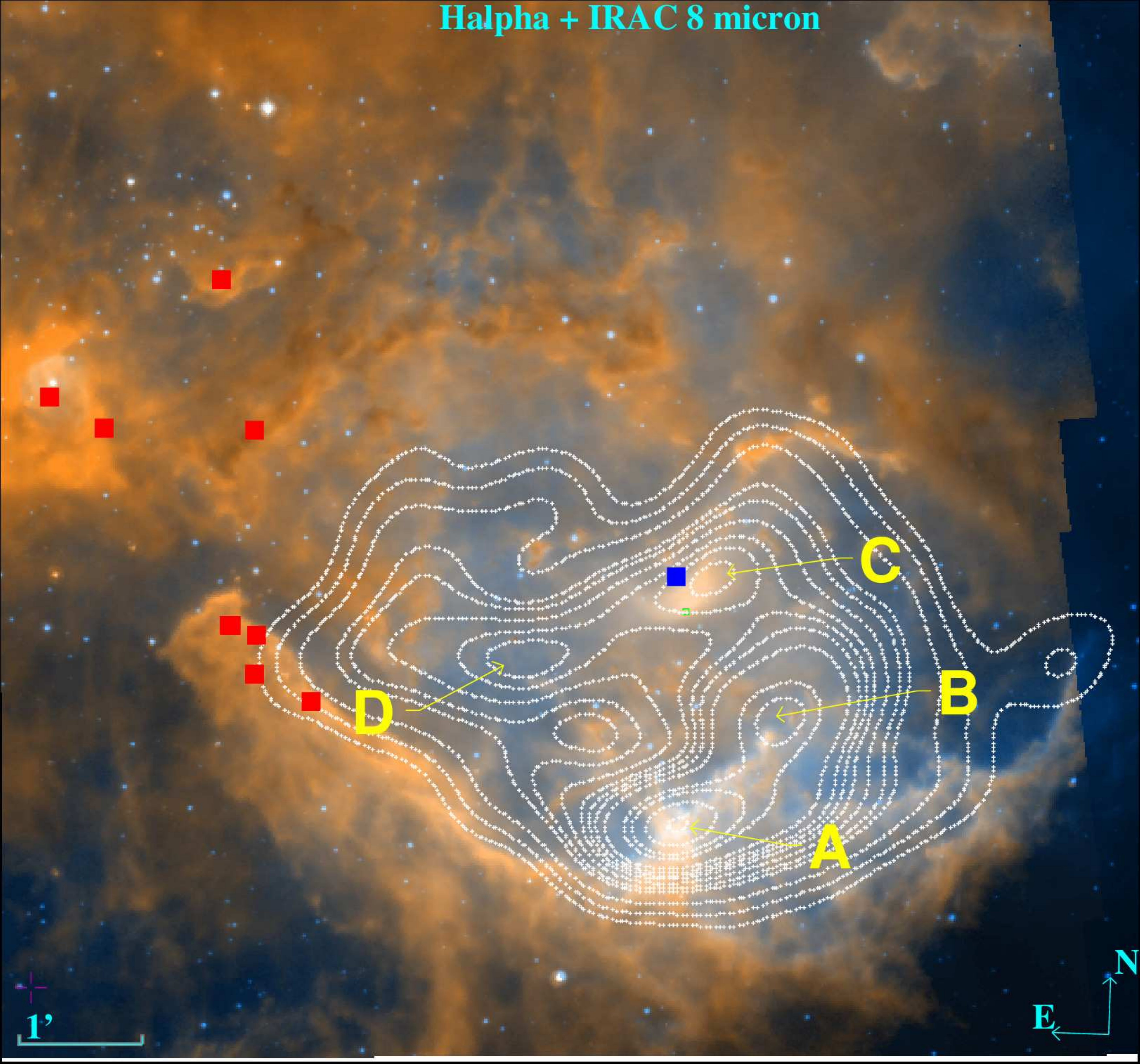}
\includegraphics[trim=0.0cm 0.3cm 0.0cm 0.0cm, clip, width=8.4cm]{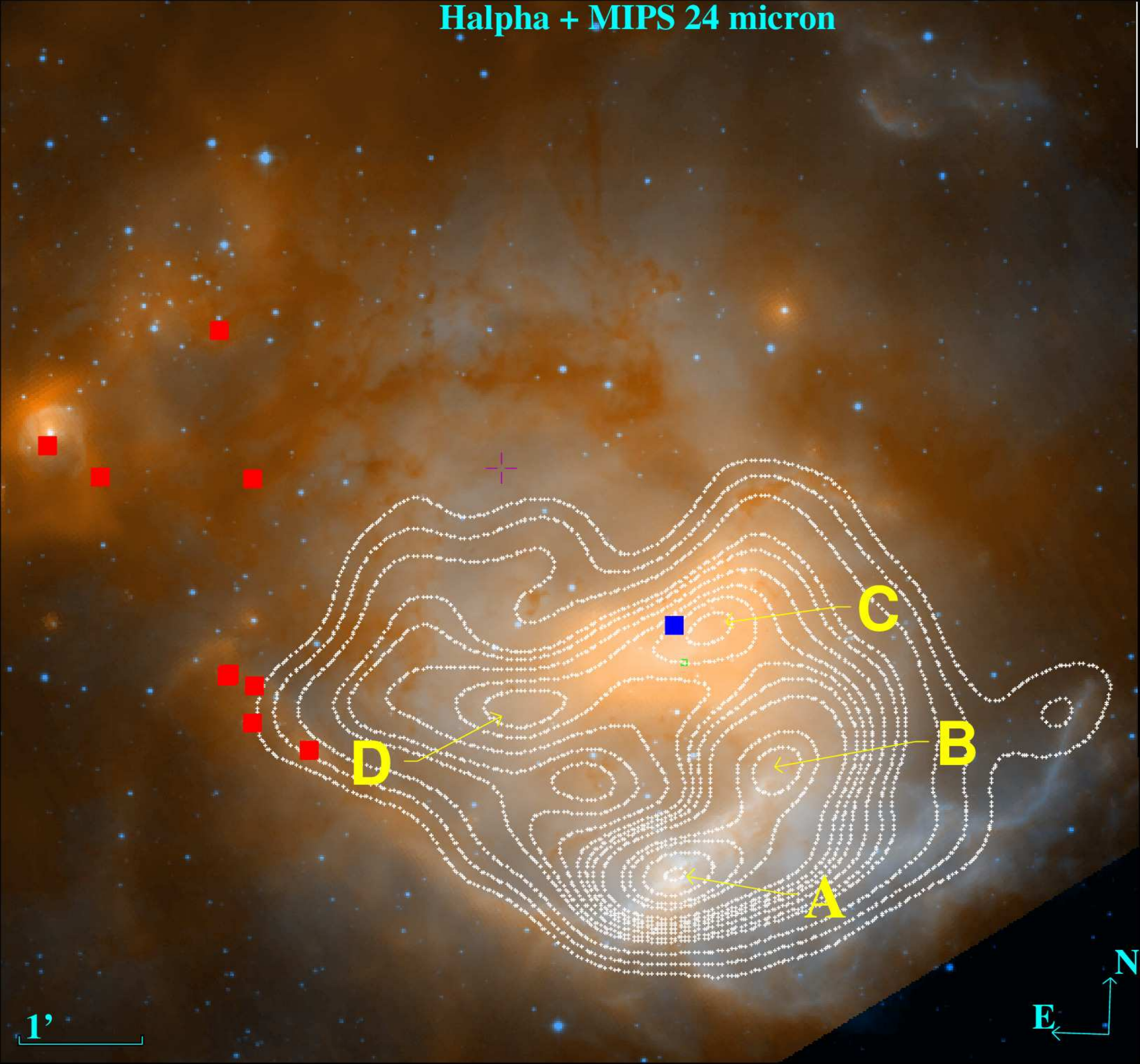}
}
\caption{Left panel: {\it Spitzer} 8 \mum image (red) superimposed on \halpha image (blue) taken with WFI camera at the 2.2-m MPG/ESO Telescope. Right panel: {\it Spitzer} 24 \mum image (red) superimposed on a WFI \halpha image (blue). The GMRT radio contours at 1280 MHz are displayed on both the images. The locations of Class I sources and the ionizing source are marked with solid red squares and a blue square, respectively on both the images. The field size is $\sim$9.2$\times$9.1 arcmin$^2$ and north is up, east is to the left for all the images.}\label{morp_im}
\end{figure*}

Since massive stars in SFRs significantly affect the entire region, the morphology of SFRs is an important tool to understand the star formation history of the region. 
 The left panel of Fig. \ref{morp_im} presents a colour-composite image made by using \halpha (blue) taken from \citet{Yadav2015} and IRAC ch4 (red),  whereas the right panel shows a  colour-composite made from MIPS 24 \mum (red) taken from SHA and \halpha (blue) images overlayed with the GMRT  1280 MHz contours. The solid red squares and a blue square are the locations of the Class I sources and the ionizing source, respectively. A dark lane of molecular cloud (right panel) is seen in the upper half of the \halpha image, extending from East to West in a semi-circular filamentary fashion. A bright emission in \halpha bisects this dark lane to make East and West components. It is to be noted that \halpha and ch4 emissions are remarkably anti-correlated. The IRAC ch4 emission makes a rather spherical cavity around the ionizing source which is open towards East. 

The S311 region is a rather spherical \hii region around the ionizing star and is nearly surrounded by a dust ring, as revealed by the MIR emission in the ch4 observations centered at 8 $\mu$m. 
The far-ultraviolet (FUV) radiation can escape from the \hii region and penetrate into the surface of the molecular clouds to some extent creating a photo-dissociation region (PDR) in the surface layer of the clouds \citep[see, e.g.,][]{Povich2007,Deharveng2010}. PAHs within the PDR are excited by the UV photons and re-emit their energy at MIR wavelengths, particularly between 6 and 10 $\mu$m.  The ch4 includes several discrete PAH emission features (e.g., 6.2, 7.7, and 8.6 $\mu$m) in addition to the contribution from the thermal continuum component from hot dust \citep{Peeters2011}. The ring of PAH emission surrounds the IF, indicating the interface between the ionized and molecular gas (i.e. PDR). The absence of 8 $\mu$m emission in the interior of the \hii region is interpreted as due to the destruction of PAH molecules by intense UV radiation of the ionizing star \citep{Povich2007}. The dust ring represents a bubble with the ionization source at its center. The bubble seems to be open towards the east direction.  
Inside the bubble the region is extremely rich in structure. \citet{DeMarco2006} observed the central part of the bubble (FoV 150 $\times$ 150 arcsec$^2$) using HST ACS camera and discovered several extremely sculpted limb-brightened ridges and many small and partly limb-brightened fragments. In their study they also found few finger-like structures with tips pointing towards the ionizing source. 

The right panel of Fig. \ref{morp_im} presents the distribution of emission at 24 \mum and \halpha wavelengths. Extended 24 \mum bright emission associated with the bubbles is observed around the ionizing source.   
Spectroscopy of the \hii regions \citep{roelf1998, peet2002} shows that no strong nebular emission lines are present in the 24 \mum band; the emission is mostly from the continuum and can only be attributed to the dust. Therefore, warm dust grains must be present in the ionized region.  
Emission at 24 \mum is also observed in the 
PDR region detected in the 8.0 \mum observation which delineates the cavity. This emission is rather faint.  

The low resolution ($\sim$43 $\times$ 33 arcsec$^2$) radio continuum map from GMRT at 1280 MHz (Fig. \ref{morp_im}, right panel) shows emission from ionized gas of the nebula, which indicates  
that the ionized gas emission spatially follows the same pattern as the warm dust emission at 24 \mum. It is evident from the radio morphology that there is a sharp boundary towards the southern part of the nebula and diffuse emission extending towards the NE and NW directions. The structure of the radio continuum map and \hii region indicates that the ionized gas is possibly undergoing a champagne flow \citep[also known as a blister region;][]{tenorio1979, alloin1979, Stahler2005fost.book.....S} due to the fact that density is low towards NE and NW. On the contrary, the southern region seems to be ionization  bounded.  

The spatial distribution of youngest sources (Class I; red squares) detected in the region are also over-plotted in Fig. \ref{morp_im}. The PAH emission observed at the boundary of the \hii region is remarkably well correlated with the distribution of these candidate YSOs,  
 suggesting that a few of them are thought to have formed due to the expansion of the \hii region.

The ionizing source HD 64315 ($\sim$O6 V) seems to be on the MS with a MS life-time of $\sim$5 Myr \citep{Meynet1994A&AS..103...97M}. This suggests that the \hii region may still be under expansion. The dynamical age of S311 can be estimated by using the model by \citet{Dyson1997pism.book.....D} with an assumption that the ionized gas of pure hydrogen is at a constant temperature in an uniform medium of constant density. If the \hii region S311 is created in a homogeneous medium of typical density 10$^3$(10$^4$) cm$^{-3}$ and ionized by an O6V star which emits $N_{\rm Lyc}$ $\sim$10$^{49}$ s$^{-1}$ (log$N_{\rm Lyc}$ $\sim$49.0) effective ionizing photons per second, its Str\"{o}mgren radius ($r_s$) would have been $\sim$0.13 (0.62) pc. Since the \hii region is over-pressured compared to the surrounding neutral gas and expands, its radius $r(t)$ at time $t$ \citep{Spitzer1978} is given as
\[
r(t) = r_s\left(1+\frac{7C_{II}t}{4r_s}\right)^{4/7}
\]

The observed radius of the ionized gas from the GMRT radio contours as well as from the NVSS contours is found to be $\sim$3.15 arcmin, 
which corresponds to a physical radius of about 4.6 pc at a distance of 5.0  kpc. The expansion velocity of the ionized gas is reported of the order of 9 km s$^{-1}$ \citep{Pismis1976RMxAA...1..373P}. Assuming an expansion speed ($C_{II}$ ) of 9  km/s and typical density of the ionized gas as 10$^3$(10$^4$) cm$^{-3}$, the dynamical age ($t$) of the ionized region comes out in the range of $\sim$1$-$4  Myr. The estimated dynamical age must be considered highly uncertain because of the assumption of an uniform medium, the assumed expansion velocity etc. Since the \hii region is found to be (partially) surrounded by the neutral gas, we would expect an expanding IF to be preceded by a swept-up shell of cool gas as it erodes into a neutral cloud. This is evident from the 8 $\mu$m emission shown in Fig.  
\ref{spatial_dist}.  

 All the Class I and a few of the Class II candidate YSOs are found to be associated with the PAH ring as revealed by the MIR emission in the  {\it Spitzer}  IRAC 8 $\mu$m observations (cf. Fig. \ref{spatial_dist}), suggesting triggered star formation around the region. 
We find that the candidate YSOs do not show any aligned distribution as expected in the case of RDI, so RDI cannot be the process for triggering of the second generation of stars in S311.  
\citet{snider2009ApJ...700..506S} preferred the SF traveling in advance of the IF as the triggering mechanism over the CC scenario because they do not see any strong evidence for regularly spaced protostars forming around the edge of the \hii region. However, they pointed out that some of the candidate YSOs in the region might have been formed by the CC process as well.  
We speculate, on the basis of spherical morphology of the ionized region, which is partially surrounded by the dust emission (see Fig. \ref{spatial_dist}),  
and the association of Class I and Class II candidate YSOs   
with the dust region, triggered star formation, possibly due to the CC process took place around the S311 region. 

\section{Summary \& Conclusions}
\label{concl}
 We have performed multiwavelength investigation of young stellar population and star formation activities in the \hii region S311. The main results are summarized as follows: 
\begin{enumerate}
\item The 8 $\mu$m image displays an approximately spherical cavity which is possibly created due to the expansion of the \hii region. This cavity seems to be open towards the east and west directions. 
\item We have identified 125 likely candidate YSOs using IRAC and NIR TCDs which include 8 Class I, and 117 Class II sources. The ages and masses of the candidate YSOs are estimated by using the SED modeling and $V$/\vi CMD. The ages and masses are found to be in the range $\sim$0.1$-$5 Myr and $\sim$0.3$-$6 \msun, respectively. All the Class II candidate YSOs are found to be located at the distance of the S311 region, i.e. at $\sim$ 5 kpc. 
\item The extinction in the region is found to be variable with large extinction regions distributed near the periphery of the \hii region. 
\item  All the Class I and a few  of the Class II candidate YSOs are found to be associated with 8 $\mu$m emission at the periphery of the \hii region. The spatial distribution of the candidate YSOs displays a strong clustering at the NE border of the complex. 
\item The GMRT radio continuum observations at 1280 MHz detect four clumps/compact \hii regions which could have been assembled during the expansion of the \hii region.
\item The spatial distribution of candidate YSOs 
along the periphery of the \hii region, the ages of the candidate YSOs, the dynamical age of the region and the age of exciting source, suggest that the formation of candidate YSOs possibly have been triggered by the expansion of the \hii region. However, molecular line observations, detailed velocity and age measurements of the stars in the region are necessary to say anything conclusively about the star formation scenario. 
\end{enumerate}

\section*{Acknowledgments}
The authors are thankful to the anonymous referee for a thorough and critical reading. The useful comments and suggestions made by the referee improved the content of the paper. The authors also are thankful to the staff of Victor Blanco 4-m Telescope operated by the Cerro Tololo Inter-American Observatory, NOAO and the staff of GMRT operated by the National Centre for Radio Astrophysics of the Tata Institute of Fundamental Research for their assistance and support during the observations. Part of the work has been carried out at Tata Institute of Fundamental Research, Mumbai, India. This research has made use of the Simbad and Vizier databases maintained at CDS, Strasbourg, France; NASA's Astrophysics Data System Bibliographic Services; the WEBDA database operated at the Department of Theoretical Physics and Astrophysics of the Masaryk University; and the NASA/IPAC Infrared Science Archive operated by the Jet Propulsion Laboratory, California Institute of Technology. We also acknowledge the STARLINK package for command-line processing of tabular data. The STARLINK project was a long running UK project supporting astronomical data processing.  

\appendix

\begin{table*}
\caption{A sample table of $JHK_s$ photometric data of stars in the S311 region. The complete table is available in an electronic form only.}\label{phot_data}
\begin{tabular}{cccccccccc}
\hline
   \ra       &   \dec & \jh & $\sigma_{(J-H)}$ & \hk & $\sigma_{(H-K)}$ & \jk & $\sigma_{(J-K_s)}$ & \ks &  $\sigma_{K_s}$ \\
   (degree) & (degree)   &  (mag)  & (mag)          & (mag) &  (mag)        &  (mag) &  (mag)   & (mag) & (mag) \\  
\hline
 118.146242& -26.366478& 0.504 &  0.023& 0.112 &  0.027& 0.62 &  0.027& 7.419 & 0.027 \\    
 118.102319& -26.483837& 0.535 &  0.023& 0.057 &  0.04 & 0.598&  0.04 & 8.334 & 0.04  \\    
 118.084529& -26.429699& 0.096 &  0.021& 0.026 &  0.024& 0.121&  0.024& 8.514 & 0.024 \\    
 118.046312& -26.35013 & 0.358 &  0.023& 0.082 &  0.019& 0.442&  0.019& 9.206 & 0.019 \\    
 118.166904& -26.376333& 0.746 &  0.024& 0.184 &  0.021& 0.938&  0.021& 9.219 & 0.021 \\    

\hline

\end{tabular} 
\end{table*}

\begin{sidewaystable}[ht]
{\tiny \tiny
\begin{tabular}{p{0.05cm}p{0.8cm}p{1.2cm}p{1.4cm}p{1.4cm}p{1.4cm}p{1.4cm}p{1.4cm}p{1.4cm}p{1.4cm}p{1.4cm}p{1.4cm}p{1.4cm}p{1.4cm}p{0.4cm}}
\hline
ID & \ra & \dec & {$U \pm \sigma{_U}$} & {$B \pm \sigma{_B}$} & {$V \pm \sigma{_V}$} & {$I \pm \sigma{_I}$} & {$J \pm \sigma{_J}$} & {$H \pm \sigma{_H}$} & {$K \pm \sigma{_K}$} & {$[3.6] \pm \sigma{_{[3.6]}}$} & {$[3.6] \pm \sigma{_{[3.6]}}$} & {$[3.6] \pm \sigma{_{[3.6]}}$} & {$[3.6] \pm \sigma{_{[3.6]}}$} & Class \\
   &  (deg)    &     (deg)  &          (mag)    &            (mag)    &             (mag)   &               (mag)    &            (mag)     &            (mag)     &            (mag)      &           (mag)    &             (mag)    &             (mag)    &              (mag)       &  \\
\hline
 1  & 118.151306&  -26.379614&  16.518$\pm$0.017&  15.884$\pm$0.011&  15.198$\pm$0.034&  13.906$\pm$0.024&  12.732$\pm$0.044&  12.226$\pm$0.050&  11.877$\pm$0.033&  11.075$\pm$0.002&  10.684$\pm$0.002&  10.341$\pm$0.004&   9.916$\pm$0.005 &  II \\
 2  & 118.131844&  -26.368668&  18.019$\pm$0.026&  17.580$\pm$0.019&  16.771$\pm$0.041&  15.144$\pm$0.030&  13.513$\pm$0.030&  12.970$\pm$0.030&  12.598$\pm$0.026&  12.129$\pm$0.003&  11.925$\pm$0.003&  11.475$\pm$0.009&  10.795$\pm$0.009 &  II \\
 3  & 118.178398&  -26.429420&  18.315$\pm$0.027&  17.625$\pm$0.005&  16.785$\pm$0.008&  15.639$\pm$0.004&  13.644$\pm$0.043&  12.943$\pm$0.022&  12.075$\pm$0.012&  11.099$\pm$0.002&  10.638$\pm$0.002&  10.168$\pm$0.005&   8.905$\pm$0.003 &  II \\
 4  & 118.139557&  -26.446766&   ...  $\pm$...  &   ...  $\pm$...  &  17.014$\pm$0.005&  14.935$\pm$0.004&  12.841$\pm$0.029&  11.586$\pm$0.034&  10.557$\pm$0.026&   9.069$\pm$0.001&   8.400$\pm$0.001&   ...  $\pm$...  &   ...  $\pm$...   &   I \\
 5  & 118.129349&  -26.450253&   ...  $\pm$...  &   ...  $\pm$...  &  17.154$\pm$0.005&  14.550$\pm$0.004&  12.345$\pm$0.024&  11.261$\pm$0.022&  10.875$\pm$0.019&  10.575$\pm$0.001&  10.613$\pm$0.002&  10.069$\pm$0.004&   9.036$\pm$0.003 &  II \\
\hline
\end{tabular}
\label{yso_cat} {\bf Table 2:} A sample table of the photometric data of candidate YSOs identified by using different schemes discussed in Section \ref{ysos_iden}. The complete table is available in an electronic form only.
}
\end{sidewaystable}

\begin{table*}
\begin{minipage}{\textwidth} 
 \caption{Parameters of compact \hii regions detected from  the GMRT radio analysis.}\label{rad_par}
\scriptsize
 \begin{tabular}{ccccccccc}
 \hline
 \hline
Region & \ra & \dec &   Gaussian-fitting       & PA\footnote{Position-angle}  & Beam-deconvolved              & PA  & Peak Intensity  & Integral Intensity  \\
       &(deg)&(deg) &    -size (arcsec$^2$)    &   (deg)                      & -source-size (arcsec$^2$)                  &     (deg)       &  (Jy/Beam)      &      (Jy)   \\
 \hline
A & 118.083250 & -26.462833 & 103.8 $\times$ 52.5 & 103.5 &  98.0 $\times$ 31.5 & 102.0 & 0.189 $\pm$ 0.004 &  0.726 $\pm$ 0.017 \\
B & 118.069292 & -26.449389 & 111.4 $\times$ 85.8 & 136.0 & 104.3 $\times$ 77.1 & 131.7 & 0.159 $\pm$ 0.003 &  1.069 $\pm$ 0.026 \\
C & 118.078167 & -26.431347 & 146.0 $\times$ 77.7 & 111.3 & 141.7 $\times$ 65.8 & 110.2 & 0.105 $\pm$ 0.003 &  0.842 $\pm$ 0.031 \\
D & 118.107542 & -26.440736 & 147.1 $\times$ 63.9 & 95.6  & 143.2 $\times$ 47.7 &  95.2 & 0.101 $\pm$ 0.003 &  0.668 $\pm$ 0.026 \\
 \hline
 \end{tabular}
\end{minipage}
 \end{table*}

\begin{table*}
\begin{minipage}{\textwidth}
 \caption{Inferred physical parameters from SED fits to 16 of 125 candidate YSOs. IDs are the same as in Table \ref{yso_cat}.}\label{phy_par}
\tiny \tiny
 \begin{tabular}{cccccccccc}
 \hline
 \hline
 ID &    \ra       & \dec       & $M_{\star}$    &   $t_\star$        &   $M_{disk}$       & $\dot{M}_{disk}$    &       $\av$   &  $\chi^2$ & Class \\  
    &     (deg)    &  (deg)     &  (\msun)       &    (10$^6$ yr)     &  (\msun)           &  (10$^{-8}$ \msun yr${-1}$)  & (mag) &  & (from Table \ref{yso_cat}) \\
\hline
1   &   118.151306 & -26.379614 &4.01 $\pm$ 0.24 & 0.97  $\pm$ 0.17   & 0.001 $\pm$ 0.007   &  0.026 $\pm$   0.228    &     1.79 $\pm$ 0.07  & 2.19 & II\\
2    &  118.131844 & -26.368668 &4.73 $\pm$ 0.02 & 0.41  $\pm$ 0.01   & 0.001 $\pm$ 0.001   &  0.020 $\pm$   0.018    &     2.13 $\pm$ 0.33  & 8.93 & II\\
3   &   118.178398 & -26.429420 &5.29 $\pm$ 1.04 & 2.90  $\pm$ 1.67   & 0.000 $\pm$ 0.002   &  0.031 $\pm$   0.552    &     3.91 $\pm$ 0.39  & 4.10 & II\\
5  &    118.129349 & -26.450253 &4.73 $\pm$ 0.93 & 0.09  $\pm$ 0.04   & 0.047 $\pm$ 0.068   &  1.056 $\pm$   1.600    &     3.10 $\pm$ 0.63  & 2.98 & II\\
10  &   118.145935 & -26.431547 &2.51 $\pm$ 0.59 & 2.18  $\pm$ 1.81   & 0.009 $\pm$ 0.016   &  0.299 $\pm$   1.013    &     2.84 $\pm$ 0.65  & 6.75 & II\\
11  &   118.131332 & -26.435246 &5.77 $\pm$ 0.76 & 0.05  $\pm$ 0.02   & 0.024 $\pm$ 0.037   &  2.022 $\pm$   2.156    &     4.31 $\pm$ 0.75  &13.56 & II\\
12  &   118.136330 & -26.419991 &4.68 $\pm$ 0.31 & 0.13  $\pm$ 0.02   & 0.011 $\pm$ 0.006   &  0.388 $\pm$   0.153    &     4.31 $\pm$ 0.61  & 1.49 & II\\
13   &  118.148087 & -26.442989 &6.24 $\pm$ 1.09 & 0.59  $\pm$ 0.85   & 0.041 $\pm$ 0.079   & 12.769 $\pm$  48.673    &     5.98 $\pm$ 1.46  & 2.47 & I\\
14   &  118.147987 & -26.418949 &2.37 $\pm$ 1.02 & 0.31  $\pm$ 0.40   & 0.024 $\pm$ 0.037   &  1.796 $\pm$   3.341    &     3.49 $\pm$ 1.00  & 1.14 & II\\
15   &  118.150787 & -26.429914 &3.52 $\pm$ 0.63 & 4.06  $\pm$ 2.26   & 0.003 $\pm$ 0.014   &  0.314 $\pm$   1.939    &     5.33 $\pm$ 0.76  & 1.58 & II\\
16   &  118.143349 & -26.442850 &3.17 $\pm$ 1.42 & 0.42  $\pm$ 0.78   & 0.024 $\pm$ 0.036   &  6.434 $\pm$  14.881    &     3.88 $\pm$ 1.47  & 1.01 & II\\
19  &   118.151421 & -26.428879 &2.43 $\pm$ 1.28 & 0.74  $\pm$ 0.65   & 0.018 $\pm$ 0.042   &  1.496 $\pm$   5.377    &     3.04 $\pm$ 0.87  & 7.85 & II\\
29   &  118.137253 & -26.422029 &2.25 $\pm$ 0.82 & 5.14  $\pm$ 3.80   & 0.008 $\pm$ 0.023   &  0.545 $\pm$   2.044    &     4.36 $\pm$ 1.59  & 0.93 & II\\
37   &  118.151184 & -26.397684 &2.17 $\pm$ 1.02 & 1.16  $\pm$ 1.50   & 0.012 $\pm$ 0.023   &  0.664 $\pm$   2.110    &     4.79 $\pm$ 2.08  & 2.56 & II\\
41  &   118.147346 & -26.432081 &2.14 $\pm$ 1.29 & 1.95  $\pm$ 2.61   & 0.009 $\pm$ 0.015   &  0.277 $\pm$   1.094    &     3.96 $\pm$ 1.37  & 6.10 & II\\
49   &  118.151756 & -26.436468 &4.01 $\pm$ 1.94 & 1.18  $\pm$ 1.87   & 0.029 $\pm$ 0.051   & 17.573 $\pm$  46.989    &     6.68 $\pm$ 2.40  & 8.61 & I\\
 \hline
 \end{tabular}
\end{minipage}
 \end{table*}

\begin{table*}
\caption{\label{age_mas} Ages and masses of the candidate YSOs from $V/(V - I)$ CMDs. IDs are the same as in Table \ref{yso_cat}. The complete table is available in an electronic form only.}
\begin{tabular}{cccccccc}
\hline
\hline
ID &     \ra     &   \dec   &   $V$       &          $ (V - I)$    &           $t_{\star}$     &      $M_{\star}$ & Class \\
   &  (deg)   &   (deg)  &   (mag)     &            (mag)       &              (10$^6$ yr)    &      (\msun) &  (from Table \ref{yso_cat})\\
\hline
 1  &  118.151306& -26.379614 &  15.198 $\pm$ 0.034  &  1.292 $\pm$ 0.024  &     1.17 $\pm$ 0.20   &    3.65 $\pm$ 0.30  & II  \\
 2  &  118.131844& -26.368668 &  16.771 $\pm$ 0.041  &  1.627 $\pm$ 0.030  &     0.53 $\pm$ 0.11   &    3.27 $\pm$ 0.08  & II  \\
 3  &  118.178398& -26.429420 &  16.785 $\pm$ 0.008  &  1.146 $\pm$ 0.004  &     6.47 $\pm$ 0.53   &    1.96 $\pm$ 0.05  & II  \\
 4  &  118.139557& -26.446766 &  17.014 $\pm$ 0.005  &  2.079 $\pm$ 0.004  &     0.10 $\pm$ 0.00   &    1.63 $\pm$ 0.05  &  I  \\
 5  &  118.129349& -26.450253 &  17.154 $\pm$ 0.005  &  2.604 $\pm$ 0.004  &     0.10 $\pm$ 0.00   &    1.47 $\pm$ 0.05  & II  \\
\hline
\end{tabular}
\end{table*}

\clearpage

\begin{figure*}
\centering
\includegraphics[trim=1.2cm 1.45cm 0.0cm 0.0cm, clip,width=14cm]{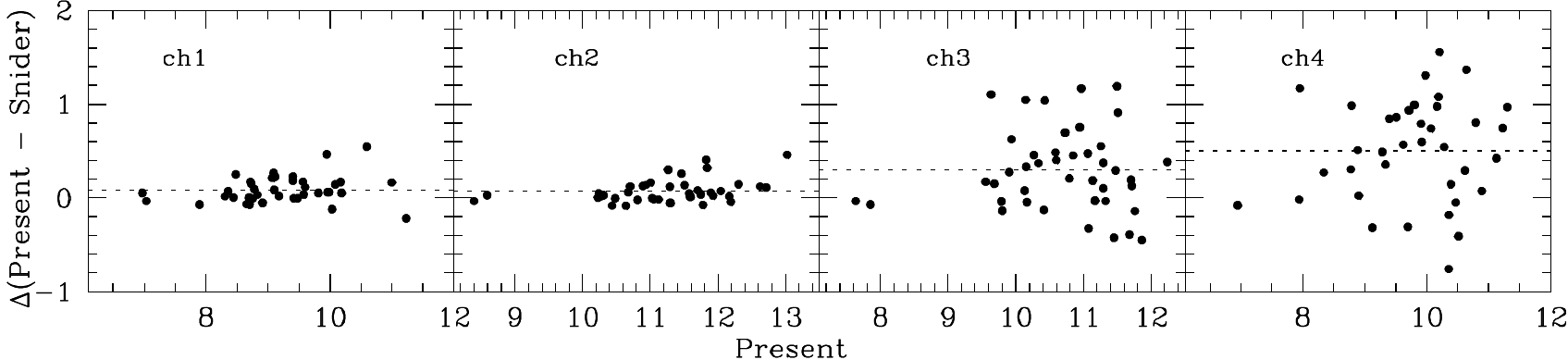}
\caption{Comparison of the present IRAC photometry with that of \citet{snider2009ApJ...700..506S} for all the four bands. The $\Delta$ represents the difference between the present and literature values. The plot axes are in magnitudes. The dashed line represent the mean of the difference between the two photometric measurements.}\label{ph_cmp} 
\end{figure*}

\begin{figure*}
\includegraphics[trim=0.0cm 0.0cm 0.0cm 0.0cm, clip, width=9cm, angle=0]{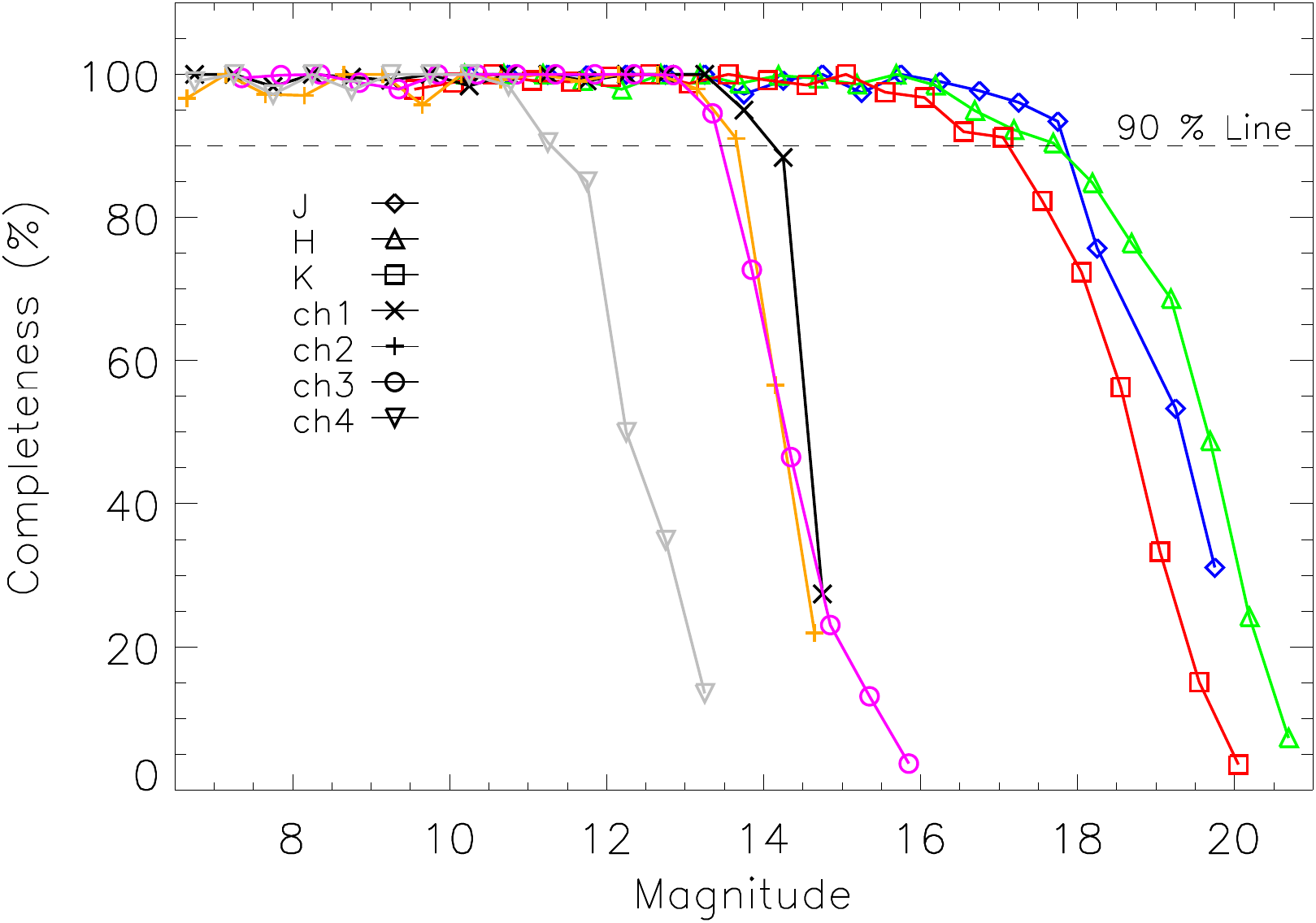}  
\caption{Completeness factors for the photometry in ISPI $J, H,$ $K_{\rm s}$ and IRAC ch1, ch2, ch3 and ch4 bands. }\label{comp_fac} 
\end{figure*}

\begin{figure*}
\centering
\vbox{
\psfrag{aaaa}{{$y$}}
\includegraphics[trim=0.0cm 0.0cm 0.0cm 0.0cm, clip, height=8cm]{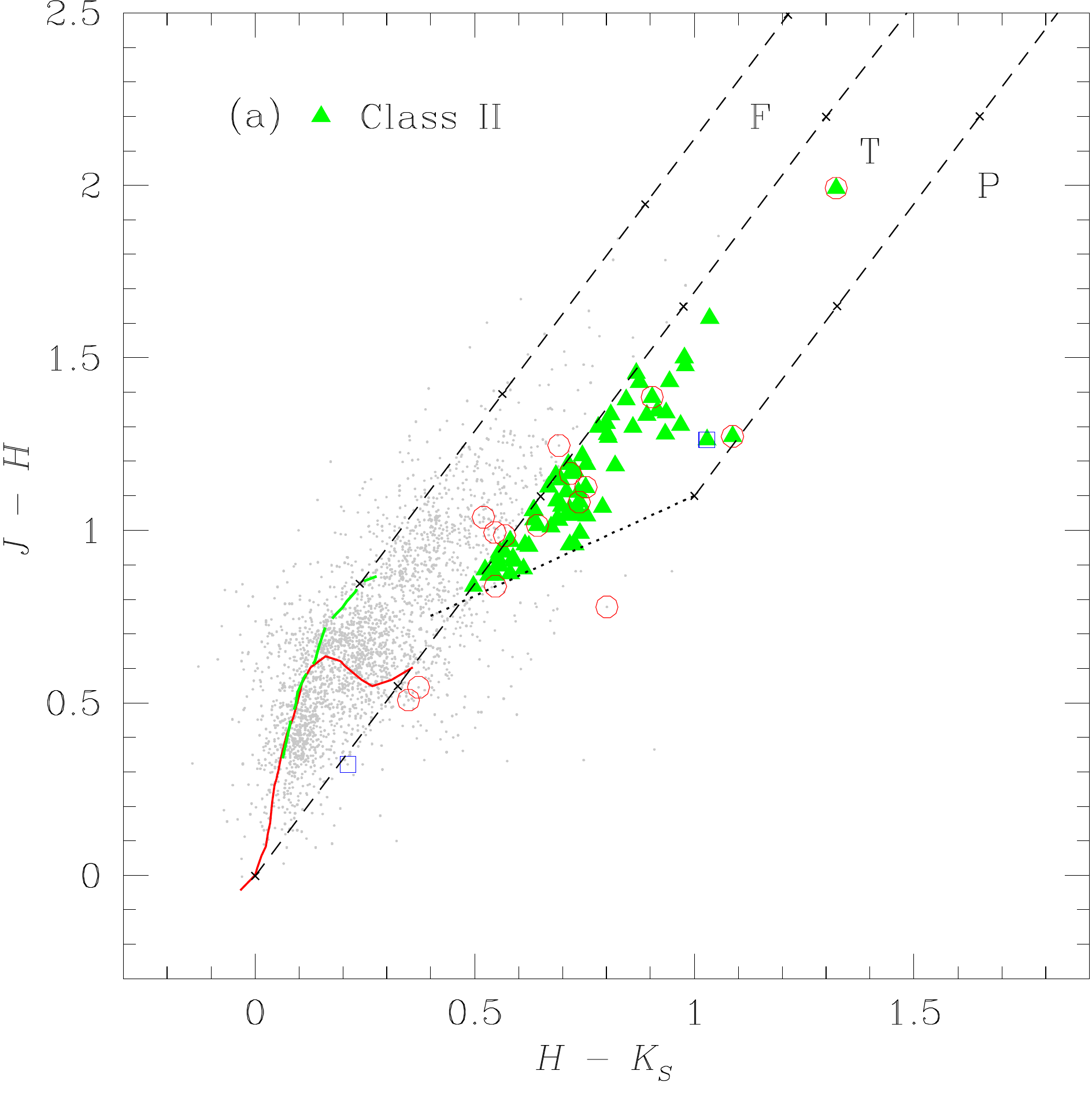}
\hbox{
\includegraphics[trim=0.0cm 0.0cm 0.0cm 0.0cm, clip, height=8cm]{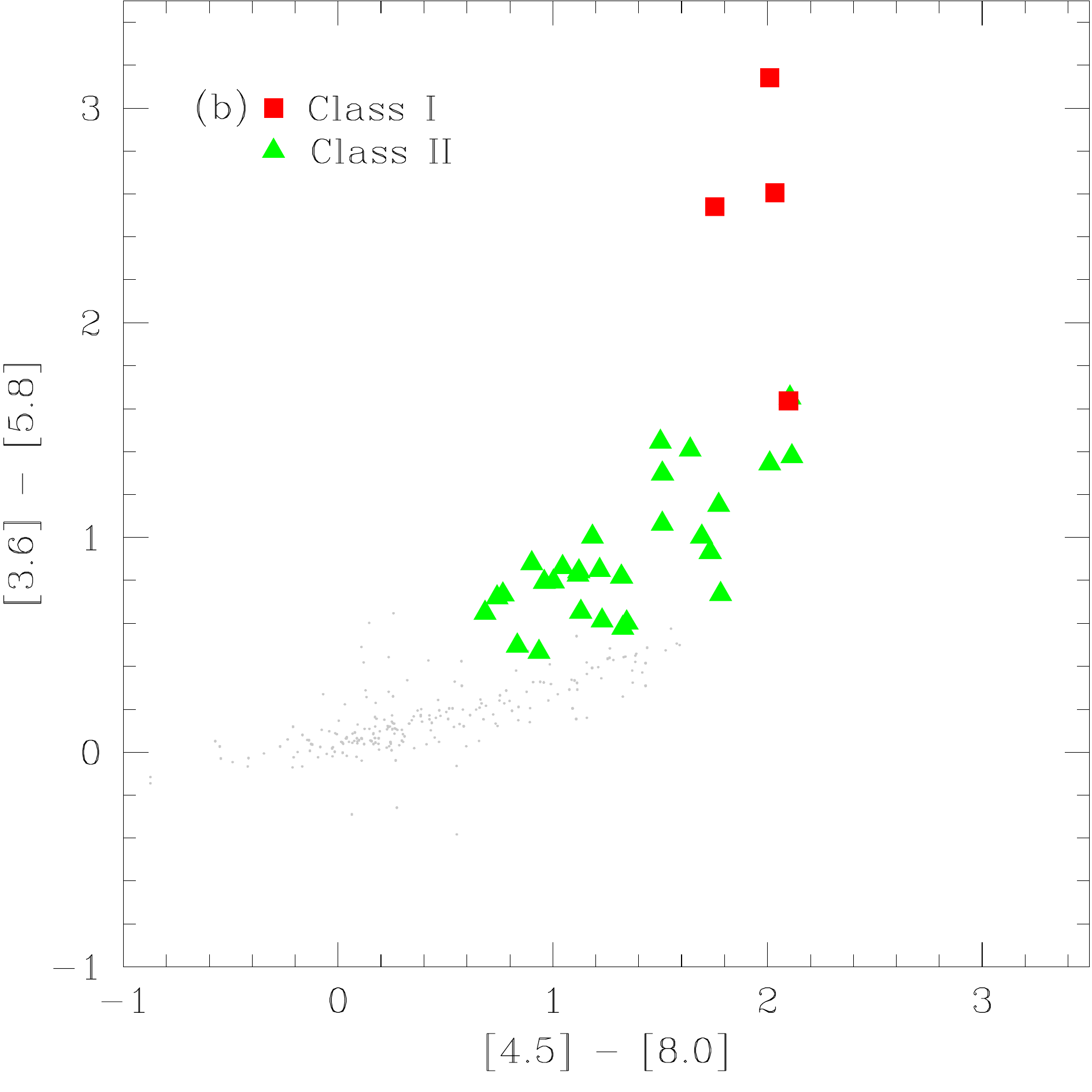}
\includegraphics[trim=0.0cm 0.0cm 0.0cm 0.0cm, clip, height=8cm]{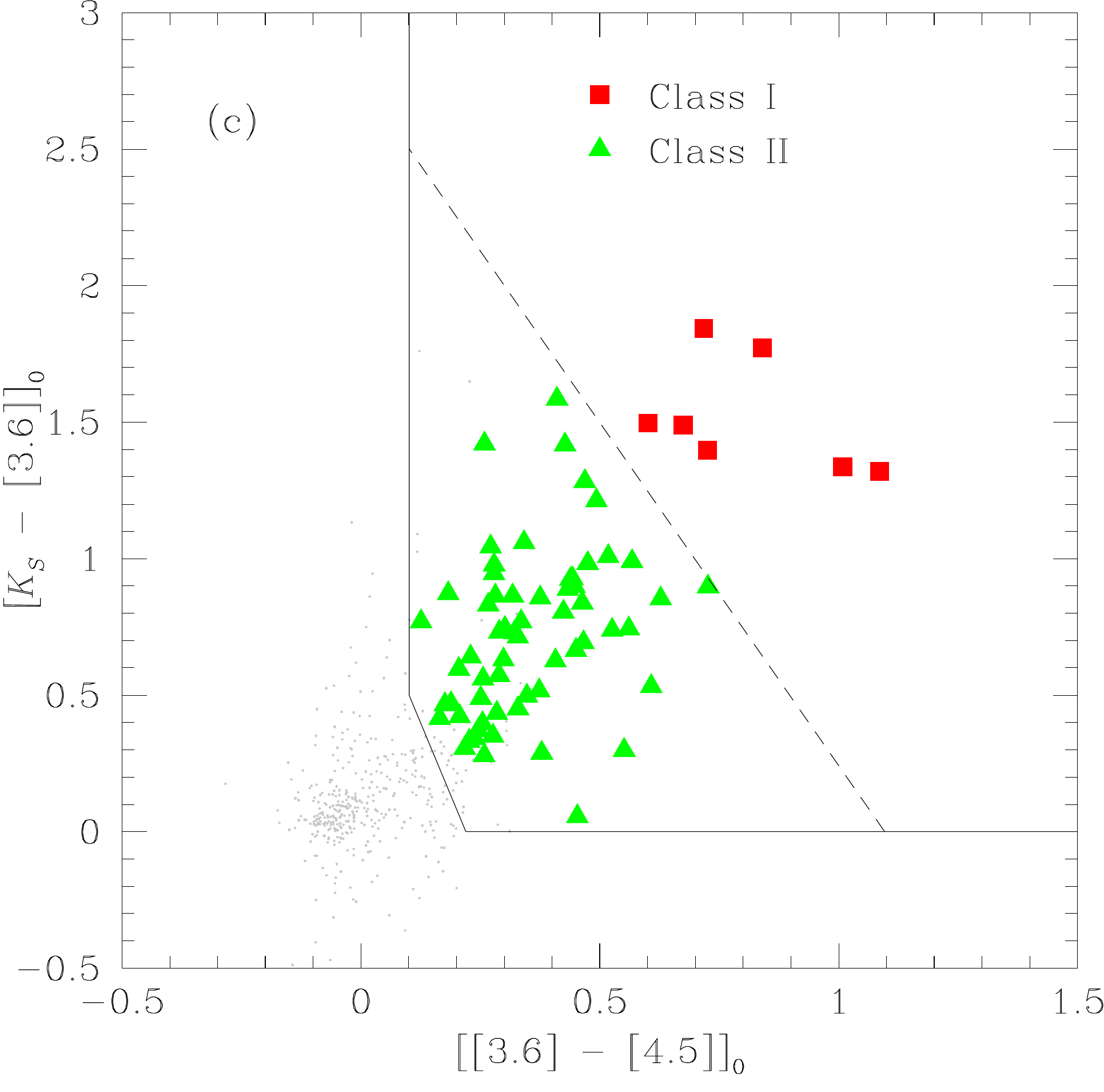}
}
}
\caption{\label{nir_tcd} (a) NIR-TCD for all the detected sources in the S311 region. The red-continuous and green-dashed curves represent the unreddened MS and giant branch \citep{Bessell1988PASP..100.1134B}, respectively. The dotted line indicates the locus of the unreddened CTTSs \citep{Meyer1997AJ....114..288M}. The parallel dashed lines are the reddening vectors drawn from the tip (spectral type M4) of the giant branch (left reddening line), from the base (spectral type A0) of the MS branch (middle reddening line), and from the tip of the intrinsic CTTS line (right reddening line) \citep{rieke1985}. The crosses on the reddening vectors show an increment of \av = 5 mag. The green triangles in the `T' region represent Class II sources. Two \halpha emission stars are shown by open squares and open circles are the $J, H,$ {\ks~} counterparts of candidate YSOs identified by using four IRAC bands (cf. Section \ref{ch1234_ysos}). (b) IRAC TCD of all the uncontaminated sources identified within the FoV of ISPI NIR observations. The candidate YSOs classified as Class I and Class II, on the basis of criteria mentioned in \citet{Gutermuth2009ApJS..184...18G}, are shown by the red squares and green triangles, respectively. (c) The dereddened ([\ks]-[3.6])$_0$ versus ([3.6]-[4.5])$_0$ TCD for all the IRAC sources having ISPI $H$, and {\ks~} counterparts after removing the contaminants. The sources in red and green colours are the candidate Class I and Class II sources, respectively.} 
\end{figure*}

\begin{figure*}
\centering
\includegraphics[trim=0cm 0.7cm 0.0cm 0.0cm, clip, width=10cm]{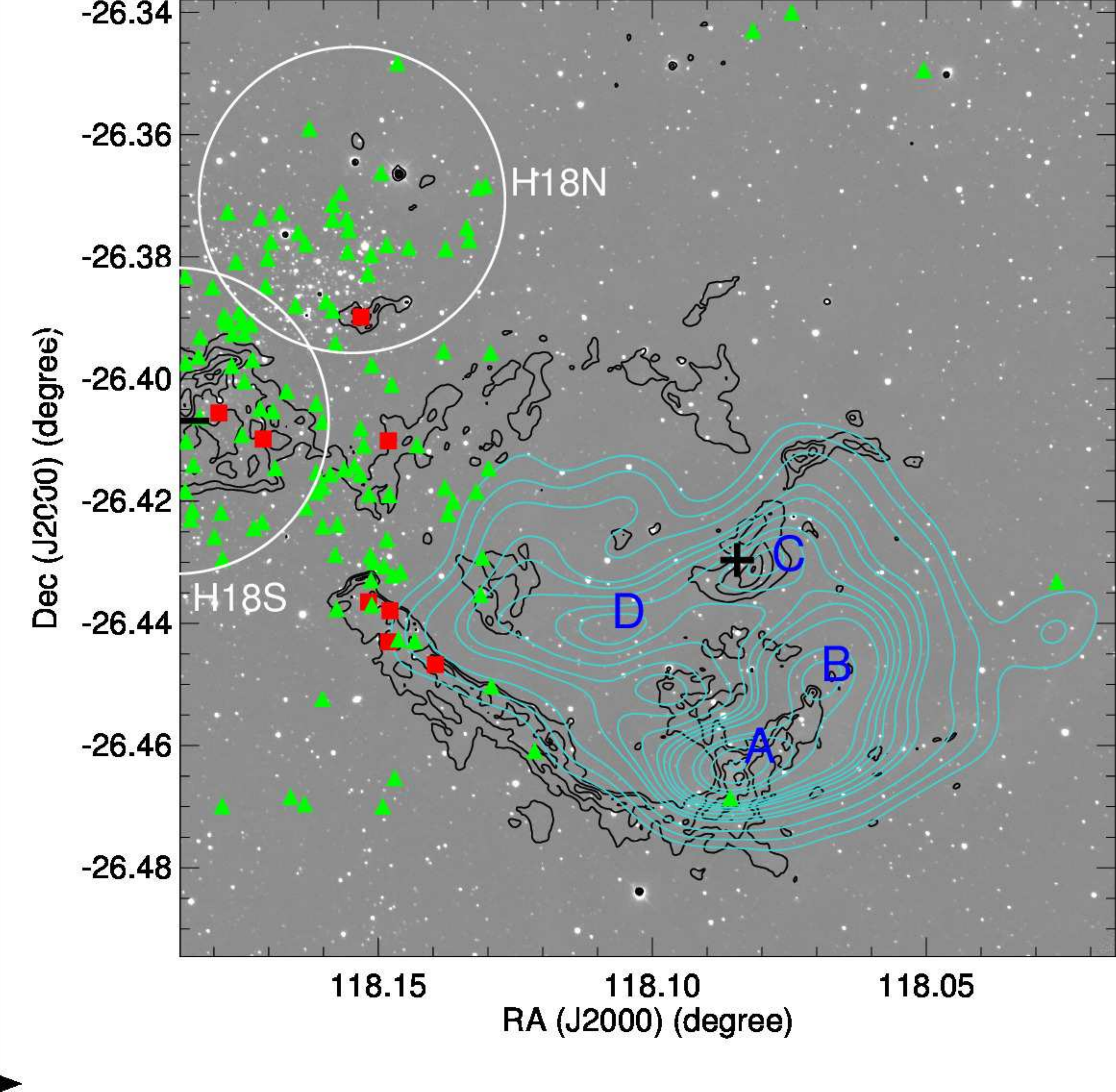}
\caption{\label{spatial_dist} Spatial distribution of candidate YSOs overlayed on the \ks-band image. The symbols are the same as in Fig. \ref{nir_tcd}. The 8 $\mu$m contours (black) and GMRT radio contours (cyan) at 1280 MHz are also over-plotted. The ionizing source (HD 64315) is shown by a {\huge +} sign. The selected sub-regions, namely H18N and H18S, are also shown. The letters A, B, C, and D denote four compact \hii regions identified from the 1280 MHz radio analysis.} 
\end{figure*}

\begin{figure*}
\centering
\includegraphics[width=8cm]{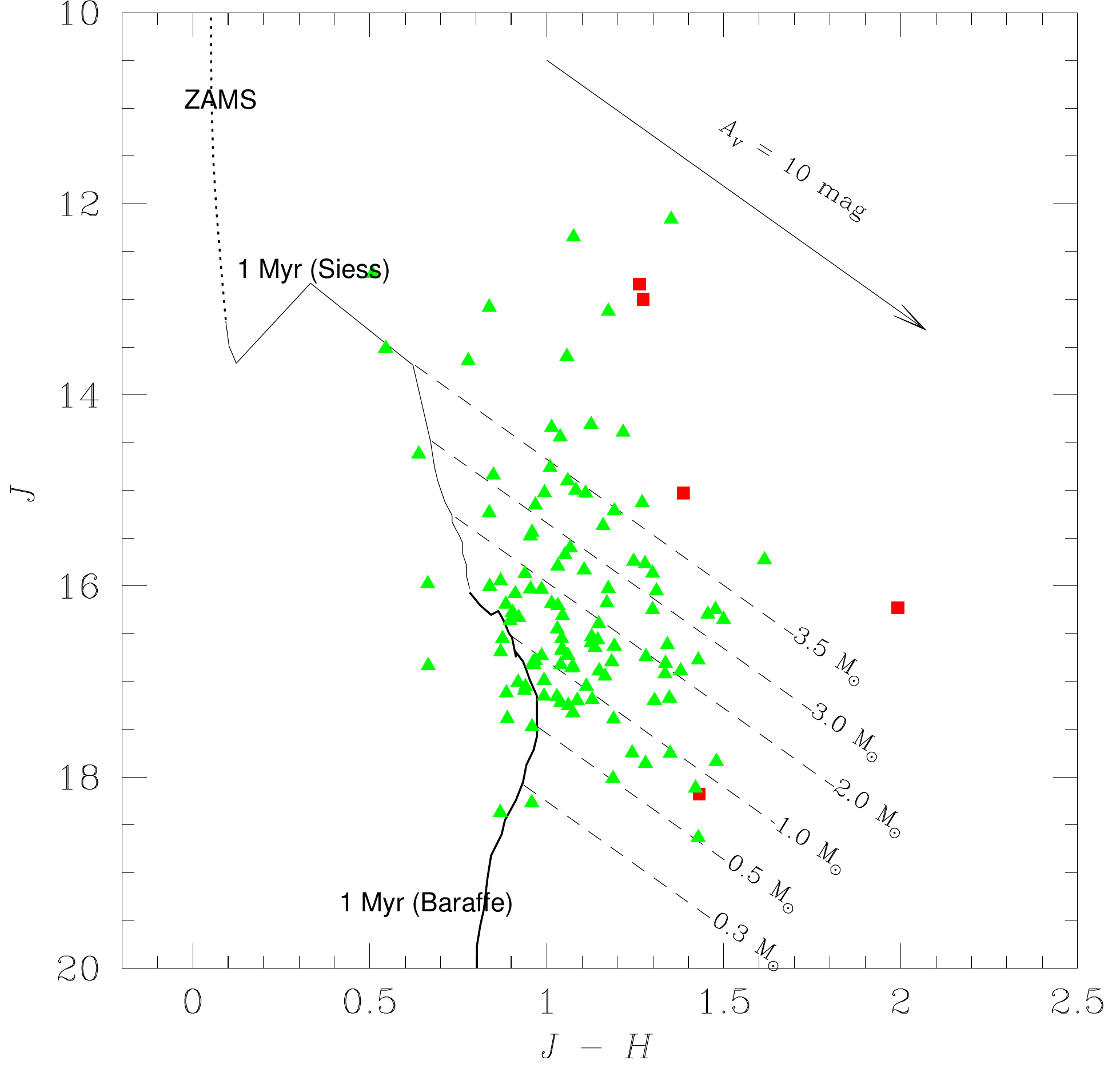}
\caption{\label{nir_cmd_ysos} $J/(J-H)$ CMD of identified candidate YSOs in S311. The symbols are the same as in Fig. \ref{nir_tcd}. The dotted curve denotes the ZAMS taken from \citet{Marigo2008A&A...482..883M}. The thin and thick continuous curves denote the loci of 1 Myr PMS star, derived from \citet{Siess2000A&A...358..593S} and \citet{Baraffe1998A&A...337..403B,Baraffe2003A&A...402..701B}, respectively. The masses range from 0.3 to 3.5 M$_{\odot}$ from bottom to top. All the isochrones are corrected for the distance and reddening of the S311 region. Dashed lines represent reddening vectors for various masses at 1 Myr.}
\end{figure*}

\label{lastpage}
\bibliography{myref}

\end{document}